\documentclass[epj]{svjour}

\usepackage{graphicx}

\usepackage[normalem]{ulem}  % \sout{old text} for strikeout
\usepackage{soul}
\usepackage{color}

\newcommand{\avep}{AV8$^{\prime}$~}
\newcommand{\avepuix}{AV8$^{\prime}$+UIX~}

\begin{document}

\title{The equation of state of neutron matter, symmetry energy, and 
neutron star structure}
\author{S. Gandolfi\inst{1} 
\and
J. Carlson\inst{1} 
\and
S. Reddy\inst{2}
\and
A. W. Steiner\inst{2}
\and
R. B. Wiringa\inst{3}}
\institute{Theoretical Division, Los Alamos National Laboratory
Los Alamos, NM 87545
\and
Institute for Nuclear Theory, University of Washington, Seattle, WA 98195
\and
Physics Division, Argonne National Laboratory, Argonne, IL 60439}

\abstract{
We review the calculation of the equation of state of pure neutron matter
using quantum Monte Carlo (QMC) methods.  QMC algorithms permit the
study of many-body nuclear systems using realistic two- and three-body
forces in a nonperturbative framework.  We present the results
for the equation of state of neutron matter, and focus on the role
of three-neutron forces at supranuclear density.
We discuss the correlation between the symmetry energy, 
the neutron star radius and the symmetry energy.
We also combine QMC and theoretical models of the three-nucleon interactions, and 
recent neutron star observations to constrain the value of the symmetry energy
and its density dependence.}

\maketitle

\section{Introduction}

Knowledge of the equation of state (EoS) of pure neutron matter at and above nuclear saturation density 
is the important bridge between the symmetry energy in nuclei and neutron star properties.
Its accurate calculation presents a challenge because of the strong interactions
between neutrons and due to the large 
extrapolation in isospin necessary to use experimental constraints from nuclear structure. 
In recent years a quantum Monte Carlo (QMC) technique, called auxiliary field diffusion
Monte Carlo (AFDMC), has been developed to study large pure neutron 
systems with the same accuracy as Green's function Monte Carlo
(GFMC), another QMC technique that has been hugely successful in calculating the
ground state properties of nuclei up to 12 nucleons using realistic nuclear Hamiltonians with local two- and 
three-body forces. Given a Hamiltonian,
the many-body system is generally solved with controlled systematic errors
within $1-2$ \%.
The combined use of GFMC and AFDMC is an 
important step forward to study the neutron matter equation of state
because the calculation of light nuclei and infinite matter is now possible
following the same scheme and within the same systematic uncertainties.

Because of its important role in astrophysics, the symmetry energy has received 
considerable attention in recent years.  The symmetry energy $E_{\rm sym}$ is
given by the difference of the energy per particle in symmetric nuclear matter and pure 
neutron matter, and represents the energy cost of sustaining an isospin-asymmetry in the
homogeneous nucleonic matter. In large nuclei and dense systems the competition between 
the Coulomb energy cost of having a large number of protons (and electrons) and the energy cost 
associated with large isospin asymmetry is key to understanding the 
mechanism of stability and neutron-skin thickness of very neutron-rich nuclei in the laboratory, 
the reaction pathways and abundances of heavy neutron-rich nuclei produced during r-process 
nucleosynthesis, neutron star structure and many related phenomena occurring in neutron stars and 
supernova \cite{Steiner:2005}. 

%The inner crust of neutron stars, where the density is a fraction
%of nuclear densities, is mostly composed of neutrons surrounding
%a matter made of extremely-neutron rich nuclei that, depending
%on the density, may exhibit very different phases and properties.
%The extremely rich phase diagram of the neutron crustal matter is
%strongly related to the role of $E_{\rm sym}$.  For example it governs the
%phase-transition between the crust and the core~\cite{Newton:2011}
%and $r$-mode instability~\cite{Wen:2012,Vidana:2012}.  

In this paper we will review a recent  AFDMC calculation of the equation of state
of neutron matter using realistic models \footnote{We define a realistic potential as one that can reproduce measured nucleon-nucleon scattering phase shifts and properties of light nuclei.} of two and three nucleon interactions and discuss 
constraints to the symmetry energy obtained from neutron stars observations.

\section{Nuclear Hamiltonian}

\begin{figure*}
\begin{center}
\includegraphics[width=0.37\textwidth]{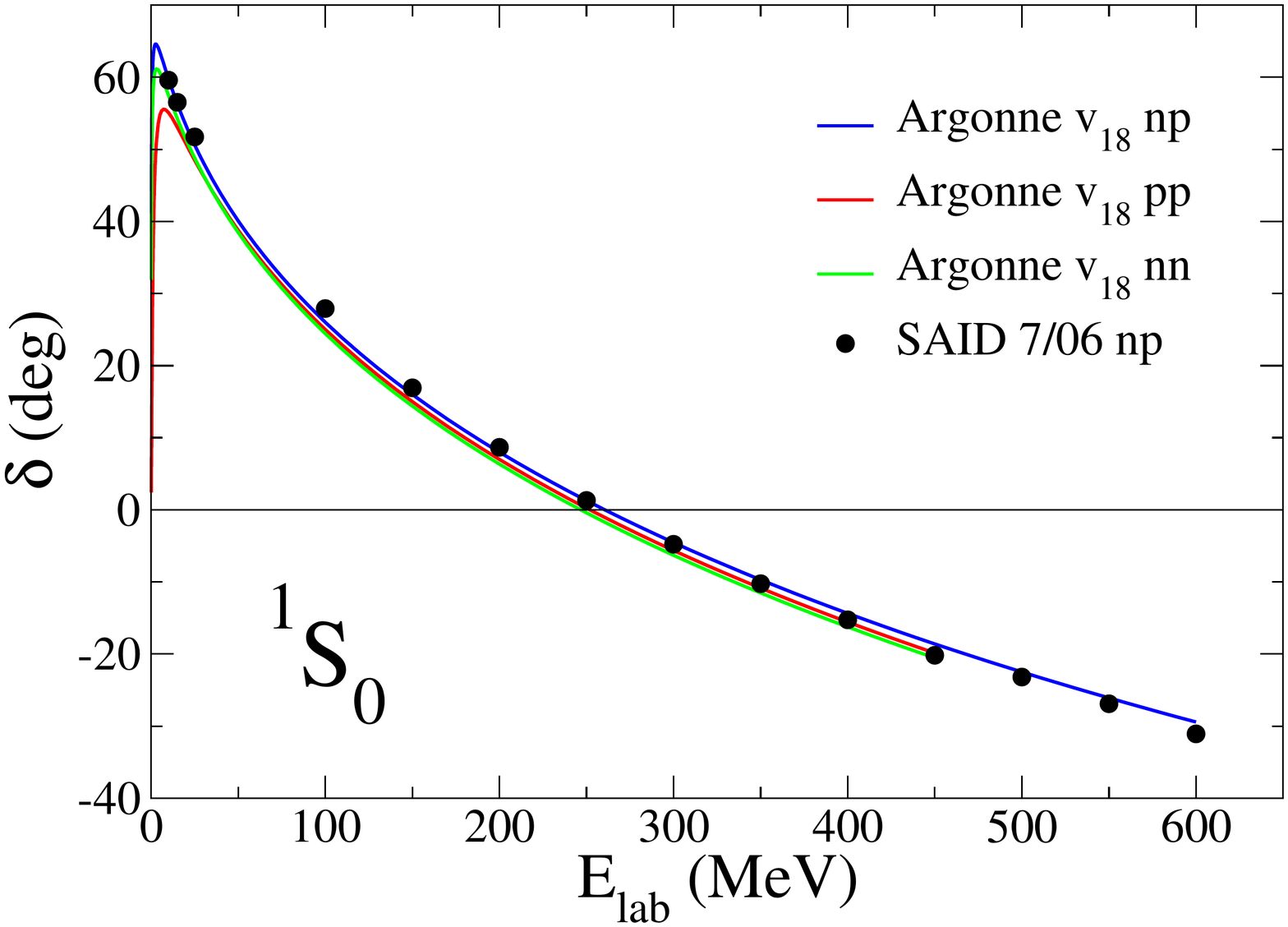}
\hspace{-1.25cm}
\includegraphics[width=0.37\textwidth]{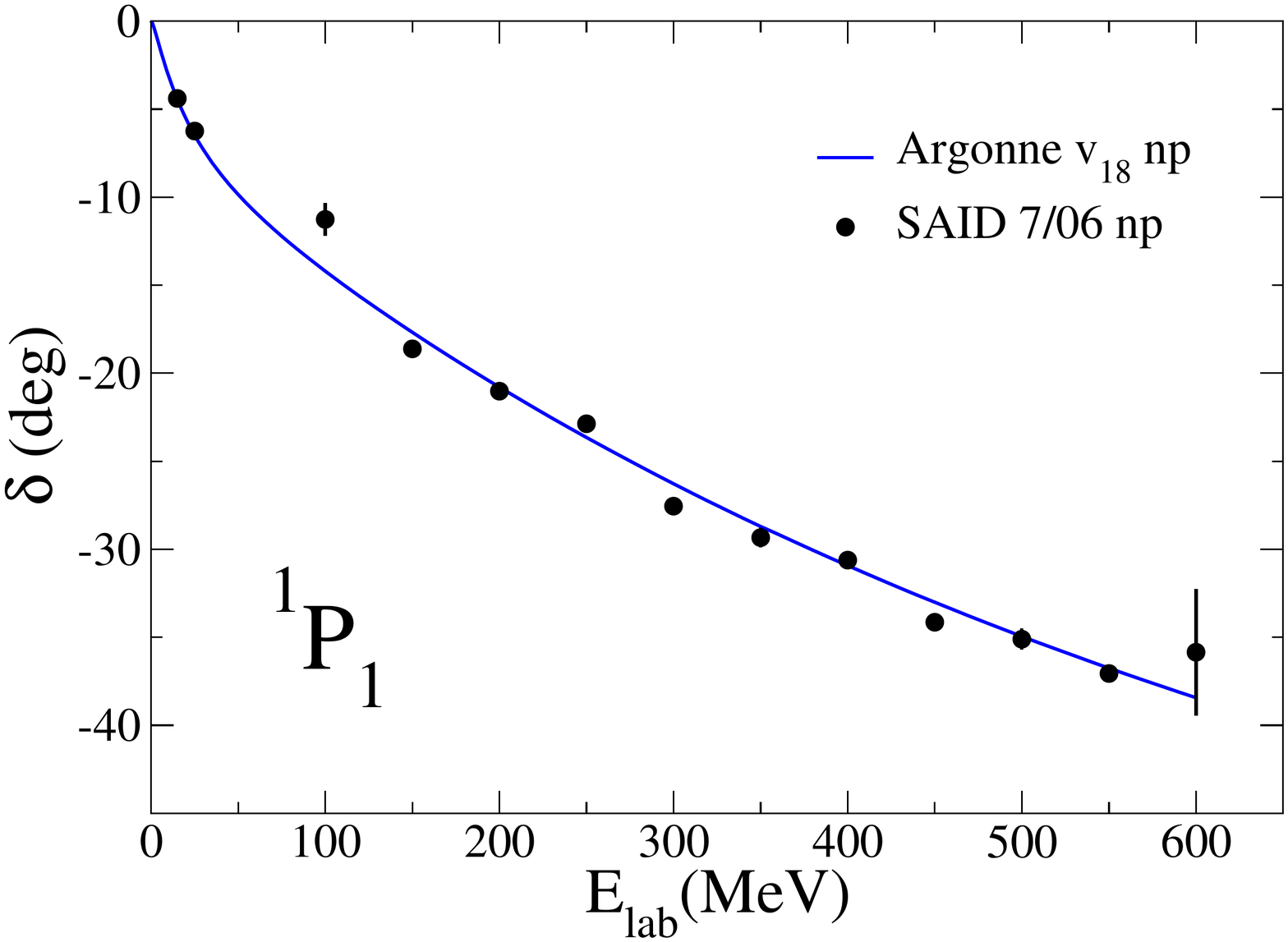}
\\
\includegraphics[width=0.37\textwidth]{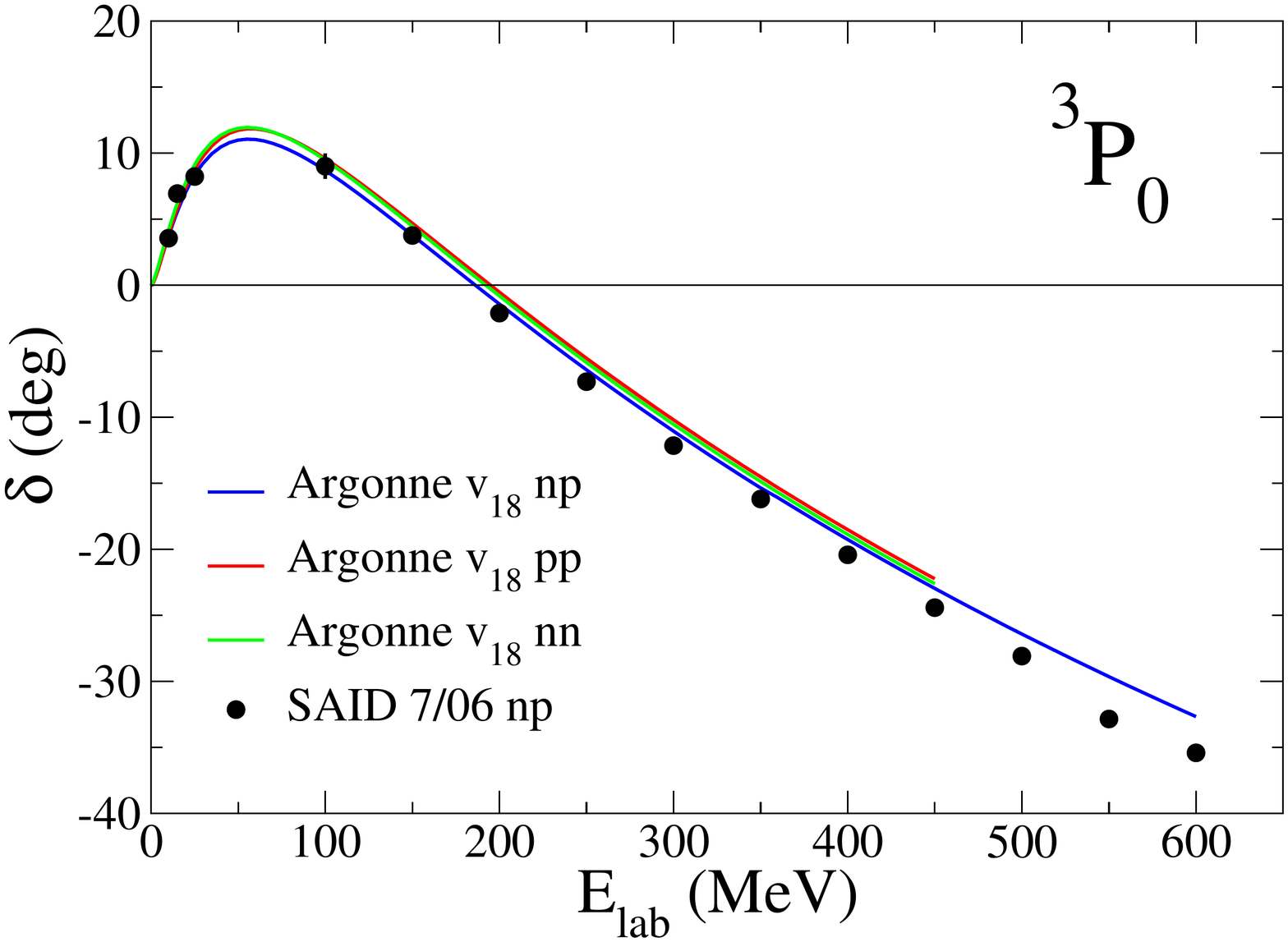}
\hspace{-1.25cm}
\includegraphics[width=0.37\textwidth]{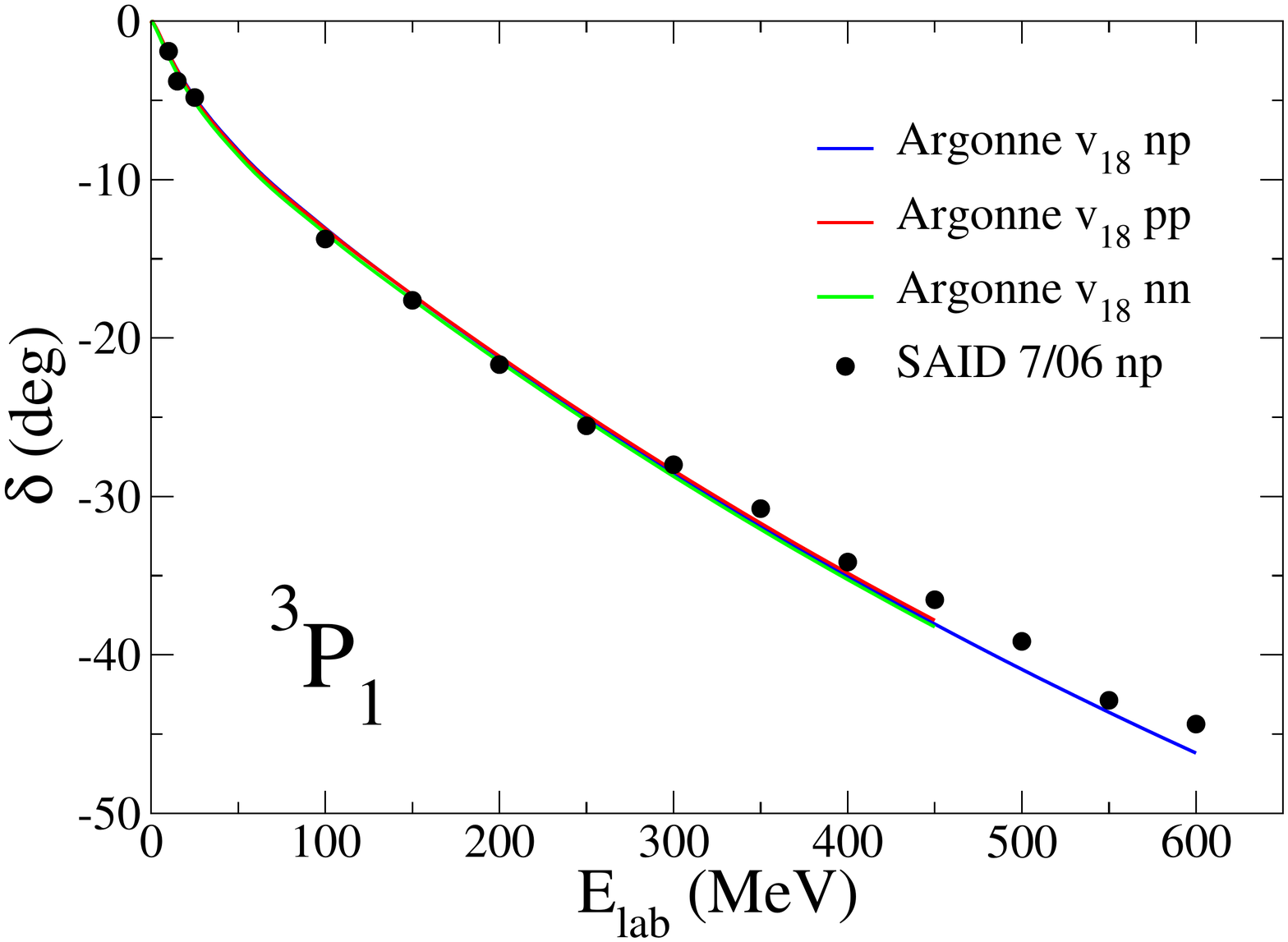}
\hspace{-1.25cm}
\includegraphics[width=0.37\textwidth]{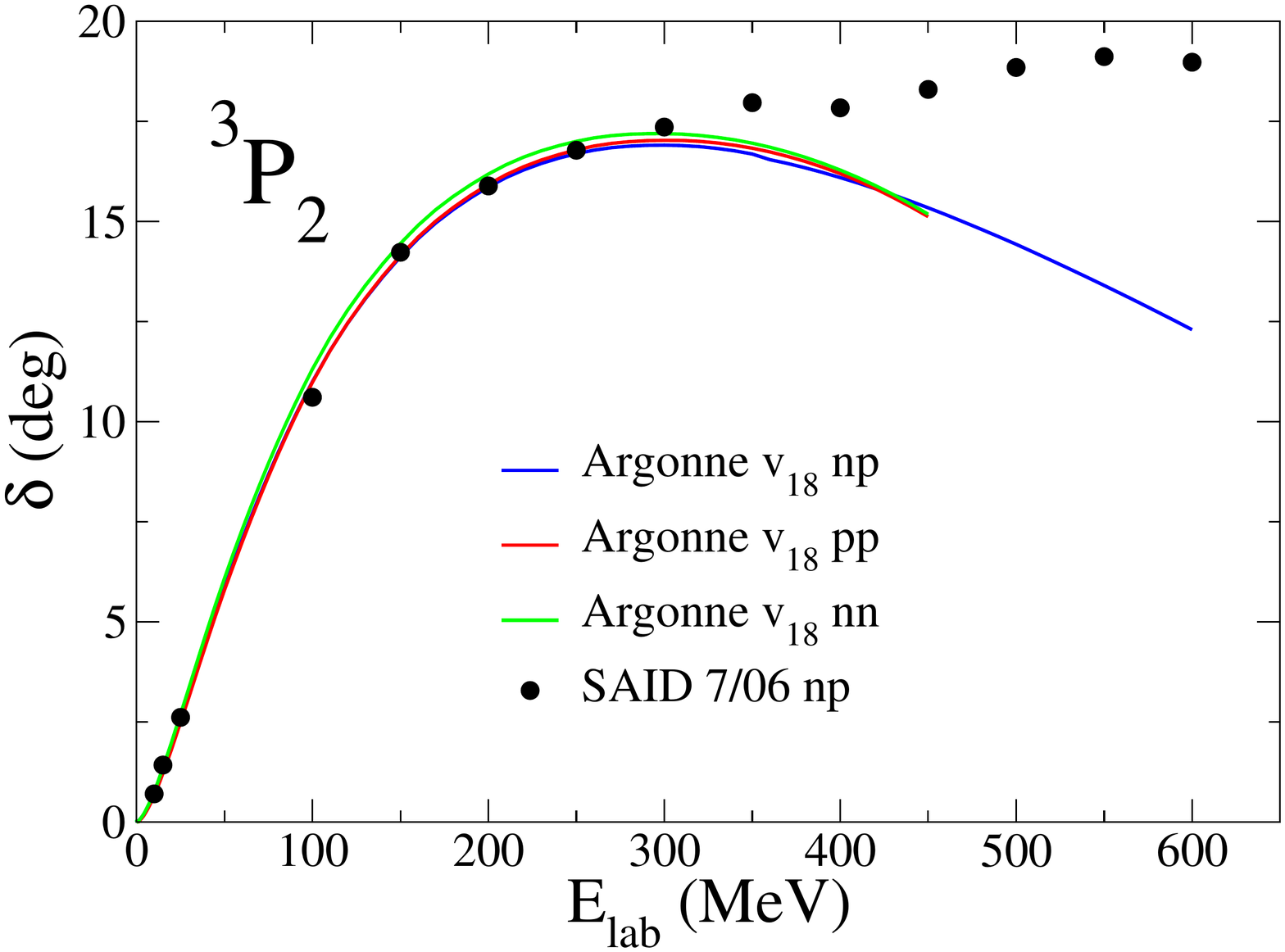}
\\
\includegraphics[width=0.37\textwidth]{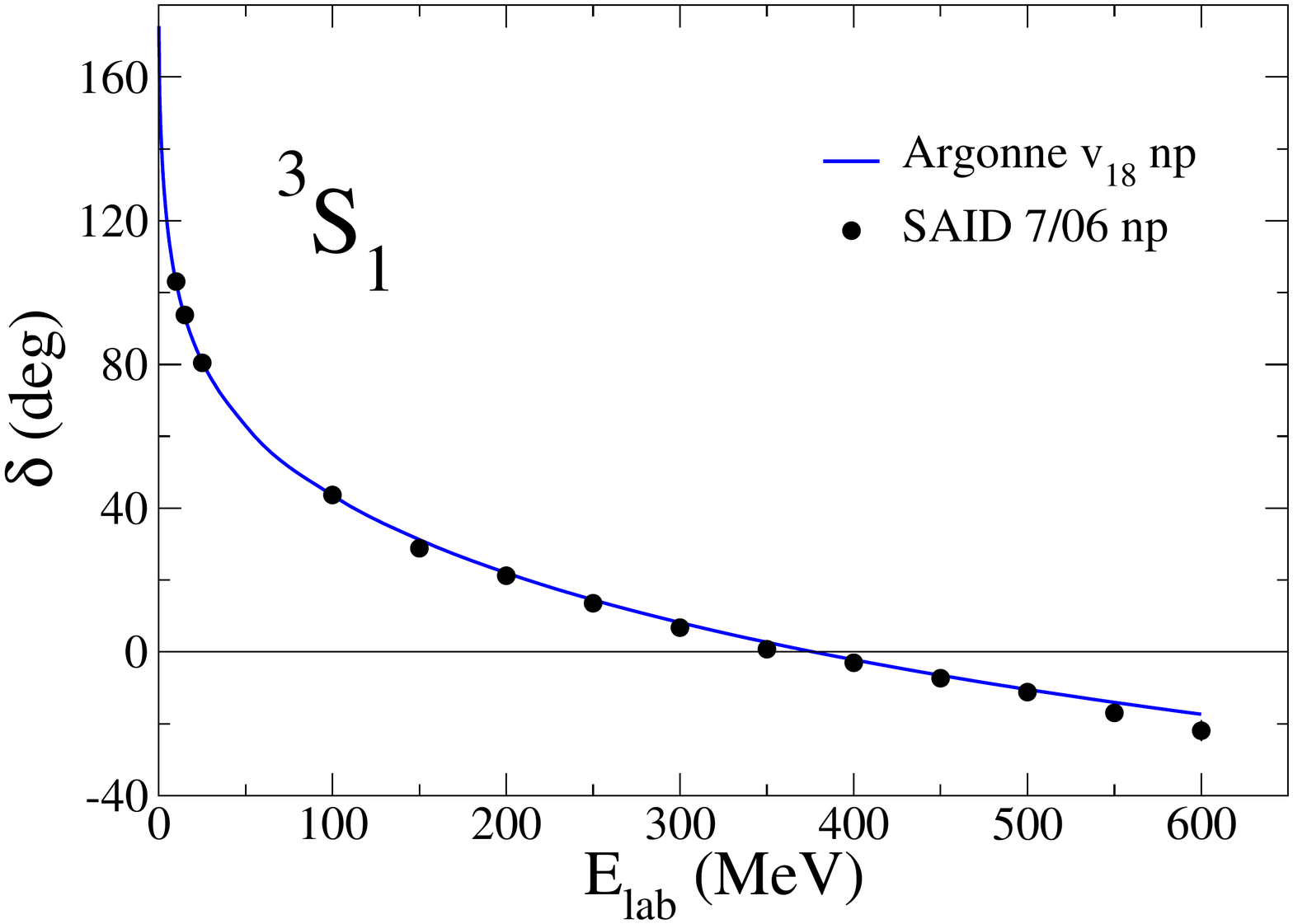}
\hspace{-1.25cm}
\includegraphics[width=0.37\textwidth]{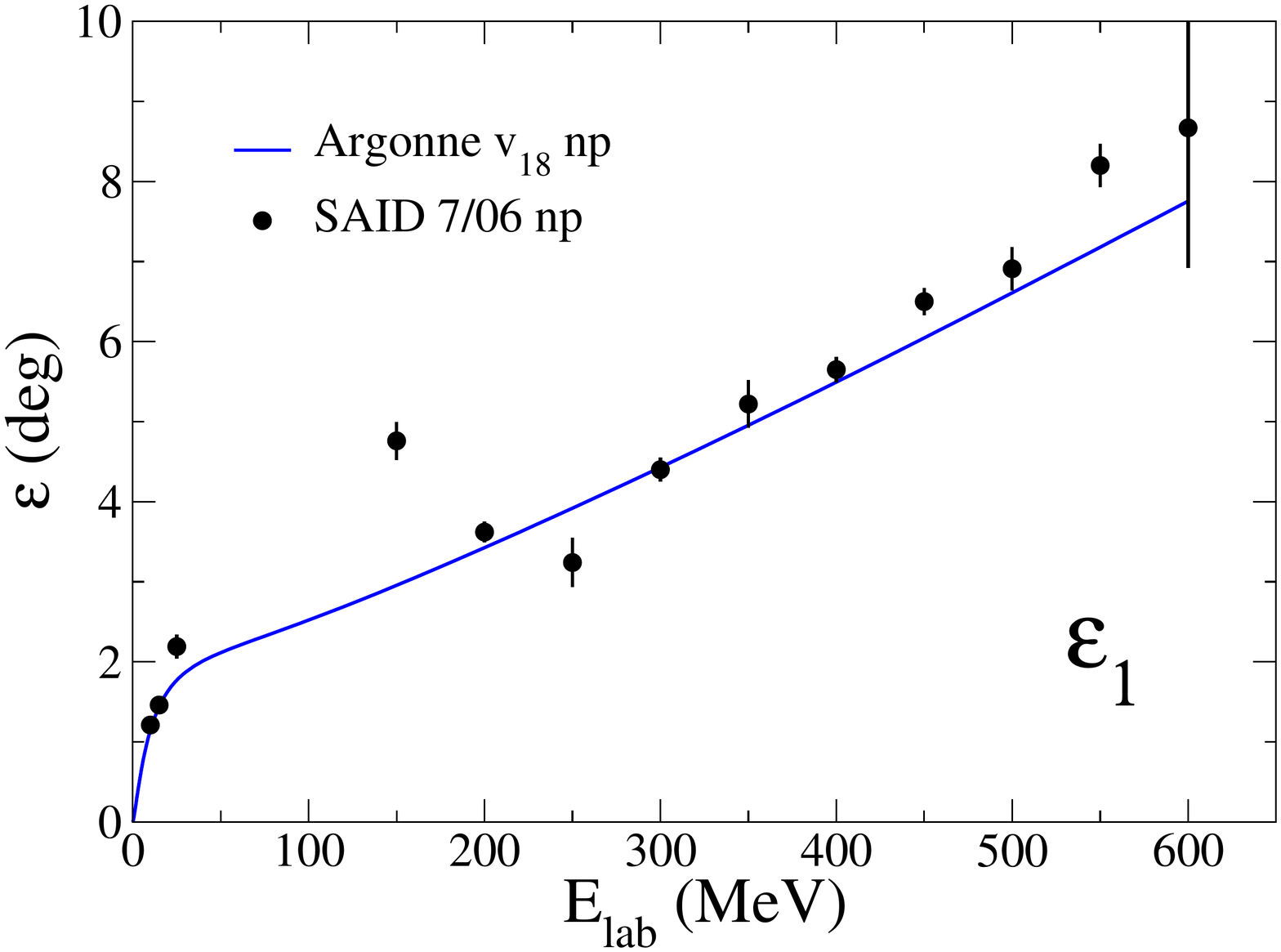}
\hspace{-1.25cm}
\includegraphics[width=0.37\textwidth]{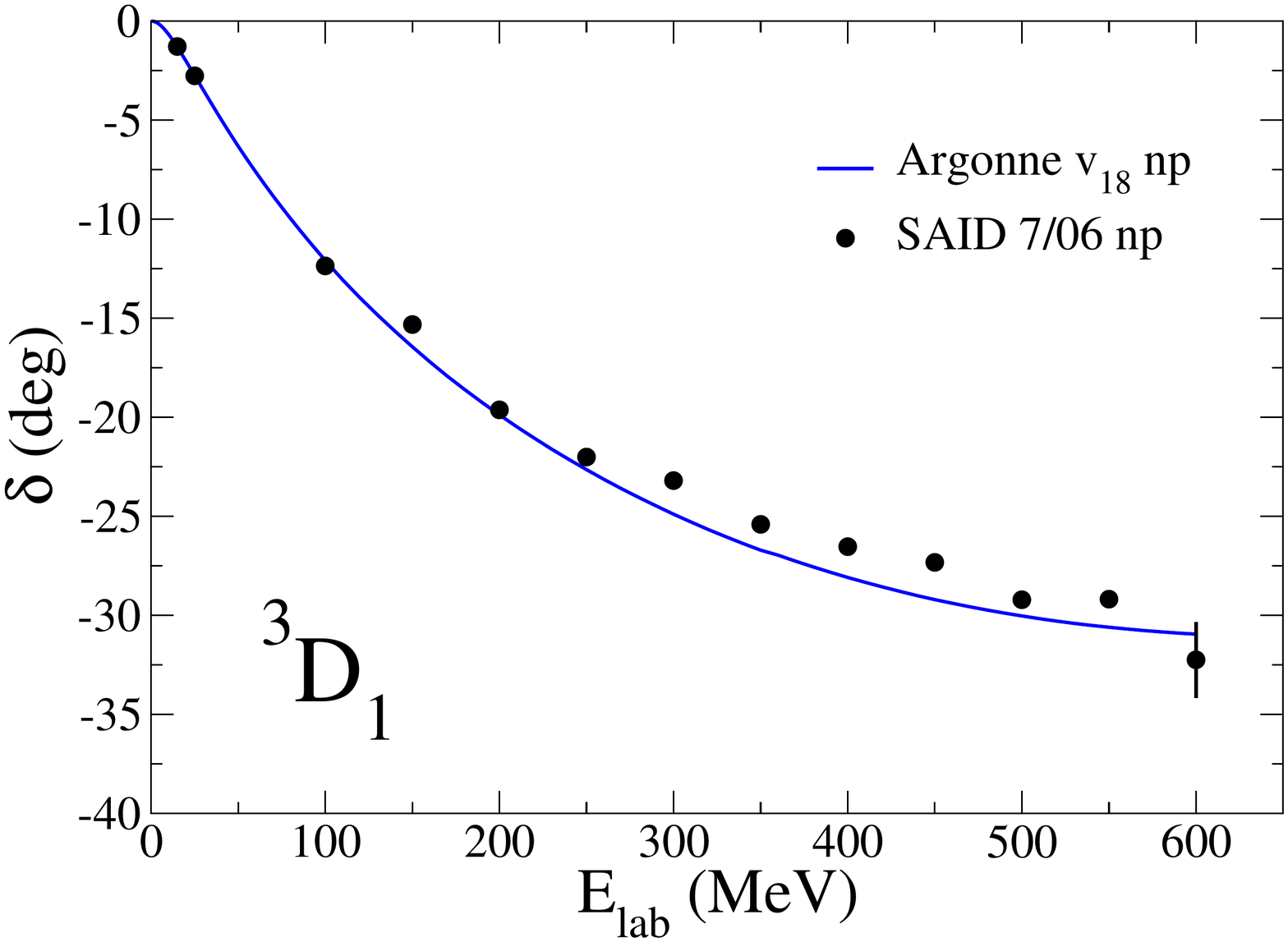}
\end{center}
\caption{Phase shifts of AV18 nucleon-nucleon potential.  
Experimental phase shifts are from the SAID Partial-Wave
Analysis Facility (gwdac.phys.gwu.edu).}
\label{fig:ps}
\end{figure*}

In our model, neutrons are non-relativistic point particles
interacting via two- and three-body forces:
\begin{eqnarray}
H = \sum_{i=1}^A\frac{p_i^2}{2m} + \sum_{i<j}v_{ij}+\sum_{i<j<k}V_{ijk} \,.
\end{eqnarray}
where $m$ is the neutron mass, and $v_{ij}$ and
$V_{ijk}$ are two and three-body potentials. Because nucleons are composite particles, 
many-body forces are natural and their relative importance increases with density. 
The two-body potential that we use is the Argonne \avep, a simplified
form of the Argonne AV18~\cite{Wiringa:1995}. This potential has been
obtained by fitting nucleon-nucleon scattering data up to laboratory energies of
350 MeV with very high precision. It has to be noted, however, that
it also reproduces many elastic phase shifts up to much higher energies, as shown in Fig.~\ref{fig:ps}. 
At saturation density, two neutrons on the Fermi surface with momenta $\pm k_F$  
($|k_F| \simeq 330$ MeV) have a center of mass
energy of $\sim$120 MeV, which would correspond to a laboratory energy of 
$\sim$240 MeV.  

It has also been shown that this 
two-nucleon potential, when combined with a model of the three-nucleon interaction, 
reproduces  low energy properties
of light nuclei with high precision~\cite{Pieper:2001b}. Four- and higher-body forces are
generally found to be negligible. In more systematic approaches, based on chiral perturbation theory, a hierarchy 
of many-body forces arises naturally  \cite{Epelbaum:2009} and explicit recent calculations of neutron matter 
in the vicinity of nuclear saturation density have shows that the contribution from four-body forces 
is very small~\cite{Kaiser:2012,Tews:2013}. 

 The two-body Argonne potential is written as a sum of operators:
\begin{eqnarray}
v_{ij} = \sum_{p=1}^{M} v_p(r_{ij}) O^{(p)}(i,j)\,,
\end{eqnarray}
where $O^{(p)}(i,j)$ are spin--isospin dependent operators.  The number
of operators $M$ characterizes the interaction; the most accurate
of them is the Argonne AV18 with M=18~\cite{Wiringa:1995}. Here we
consider a simpler form \avep, which is a reprojection of the S- and P-wave
parts of AV18 onto eight operators;
the difference between this simpler form and the full AV18 potential
can be computed perturbatively~\cite{Pudliner:1997}.  
%Most of the contribution of the NN is
%due to the one pion exchange between nucleons, but the effect of other
%mesons exchange as well as some phenomenological terms are included.
It has been shown that the difference between AV18 and \avep is less than
0.25\% in pure neutron systems~\cite{Gandolfi:2011}.

The eight $O^{(p)}(i,j)$ terms in \avep are given by:
\begin{equation}
O^{p=1,8}(i,j)=(1,\vec\sigma_i\cdot\vec\sigma_j,S_{ij},
{\vec {\rm L}}_{ij}\cdot{\vec {\rm S}}_{ij})\times(1,\vec\tau_i\cdot\vec\tau_j)\,,
\end{equation}
where the operator $S_{ij}=3\vec{\sigma}_i\cdot\hat{r}_{ij}
\vec{\sigma}_j\cdot\hat{r}_{ij}-\vec{\sigma}_i\cdot\vec{\sigma}_j$ is the
tensor operator
and $\vec {\rm L}_{ij}=-\imath\hbar\vec r_{ij}\times (\vec\nabla_i-\vec\nabla_j)/2$
and $\vec {\rm S}_{ij}=\hbar(\vec\sigma_i+\vec\sigma_j)/2$ are the relative
angular momentum and the total spin for the pair $ij$.
For neutrons $\vec\tau_i\cdot\vec\tau_j=1$, and we are left with an isoscalar
potential.
The above operators are sufficient to exactly reproduce the charge-independent
average of the AV18
phase shifts in all S- and P-waves and all deuteron properties. 
%Selected phase shifts are shown in Fig.~\ref{fig:ps} up to 600 MeV.

 The Argonne two-body force is fitted to scattering data and 
correctly gives the deuteron binding energy, but for nuclei with 3 or more 
nucleons it is not sufficient to describe the ground state of light 
nuclei. It is commonly accepted that a three-body interaction is
essential to overcome the underbinding of nuclei with more than two
nucleons.
%The importance of a three-body force is also confirmed by
%effective field theory, where its contribution appears already at the 
%N$^2$LO order~\cite{Epelbaum:2009}.
The Urbana-IX (UIX) potential has been introduced to correct this
limitation of the Argonne force.  It includes the Fujita--Miyazawa term
that describes the p-wave exchange of two pions between three nucleons, where the
intermediate state has one nucleon excited to a $\Delta$.  This term is the longest-range three-nucleon 
interaction and a very similar term also arises as the 
leading three-nucleon contribution (at N$^2$LO) in chiral perturbation theory~\cite{Epelbaum:2009}.   
In addition, we add a simple short-distance phenomenological three-nucleon interaction which is spin-isospin independent to 
soak up short distance physics ~\cite{Carlson:1981}.  The UIX force was originally proposed in combination
with the Argonne AV18 and \avep~\cite{Pudliner:1995}. Although this model of the three-nucleon force does not fully alleviate the 
underbinding problem in light nuclei, it has been extensively used to study the equation of state of nuclear and neutron
matter~\cite{Akmal:1998,Gandolfi:2009,Gandolfi:2012,Li:2008}.

Other phenomenological models of three-body force, such as the Illinois forces, have been 
obtained by fitting the binding energies of light nuclei
up to A$\le$8~\cite{Pieper:2001}. The most recent is the Illinois-7
(IL7)~\cite{Pieper:2008} which was fit to 20 states up to A=10.
It reproduces nuclear energies of 50 states (not including isobaric 
analogs) up to $A=12$ with an rms error of 600 keV. In addition to the Fujita--Miyazawa term
the Illinois models also include s-wave two-pion-exchange and also
three-pion-exchange ring diagrams. Such a general form for the Hamiltonian has been shown to describe
the spectrum of light nuclei~\cite{Pieper:2001,Pieper:2005},
n-$\alpha$ scattering~\cite{Nollett:2007}, and high-momentum components of nuclear
wave functions as observed in $(e,e^\prime p N)$ reactions~\cite{Schiavilla:2007}.

However, the three-pion rings included in the Illinois forces leads to a very strong attraction in 
pure neutron-matter~\cite{Sarsa:2003,Maris:2013} and makes the EoS too soft to be compatible with observed 
neutron star masses and radii. This is a sign of important deficiencies in our 
model of the three-nucleon forces, which will become especially important at higher density and 
larger isospin asymmetry. While chiral effective field theory ($\chi$EFT) provides a unified and systematic description of many-nucleon 
forces consistently with two-nucleon force, its regime of validity is restricted to relatively low density in the 
neutron star context. To access supranuclear densities, we currently need to rely on model Hamiltonians, and the specific 
one we discuss here must be viewed as a minimal model whose validity can in principle be tested by a combination of 
terrestrial experiments and astrophysical observations. 

Extensions of the model abound, and a large class of three-nucleon 
potentials with different spin, isospin and radial dependencies must be studied to draw definite conclusions.  In the future, it is hoped that new developments in effective field theory of nucleon-nucleon interactions~\cite{Gezerlis:2013}, and lattice QCD will both provide useful insights on the structure of three- and 
four-nucleon interactions.  Here, we attempt to the extract the error associated with poorly parametrized short-distance physics by varying the three-nucleon force.  Results obtained with relatively large variations of the coupling strengths and effective ranges nonetheless show only modest sensitivity to the details.       

%All the degrees of freedom responsible for the interaction between
%nucleons (such the $\pi$, $\rho$, $\Delta$, etc.) are integrated out
%and included in $v_{ij}$ and $V_{ijk}$.

%Another interesting class of nuclear potentials are derived within
%the chiral effective field theory. 
%However the need to include a cutoff to the nucleon's momentum
%limit their applicability to study dense neutron matter.
%Recently it has been showed that the systematic uncertainties due to
%the cutoff of these potentials can be controlled in a many-body 
%calculation~\cite{Gezerlis:2013}, but the uncertainty is already quite large at saturation
%density in neutron matter, making the calculation at large densities
%unfeasible. In addition, at N$^2$LO order the nucleon-nucleon phase
%shift that can be reproduced are only up to 150 MeV at present~\cite{Gezerlis:2013}.

\section{The AFDMC method}

QMC methods provide a very useful tool to study the
many-body ground state of strongly correlated systems without the need
of using perturbative techniques. 
The advantage of QMC methods is that the calculation can 
be done using a nuclear Hamiltonian with strong nonperturbative interactions that can describe 
high momentum observables (scattering phase shifts at large center-of-mass energy, nuclear response 
at large momentum transfer, and EoS of dense matter). 

We solve the many-body ground-state using the 
AFDMC method originally introduced by Schmidt and
Fantoni~\cite{Schmidt:1999}. The main idea of QMC methods is to evolve
a many-body wave function in imaginary time:
\begin{equation}
\Psi(\tau)=\exp[-H\tau]\Psi_v \,,
\end{equation}
where $\Psi_v$ is a variational ansatz that includes short-range two-body correlations, 
and $H$ is the Hamiltonian
of the system.  In the limit of $\tau\rightarrow\infty$, $\Psi$
approaches the ground-state of $H$.  The evolution in imaginary time
is performed by sampling configurations of the system using Monte
Carlo techniques, and expectation values are evaluated over
the sampled configurations. 

Technically, the wave function is represented as a set of configurations
called "walkers":
\begin{equation}
\langle\vec R\vert\Psi\rangle=\Psi(\vec R)=
\sum_i \langle\vec R\vert\vec R_i\rangle \langle \vec R_i\vert\Psi\rangle \,,
\end{equation}
where $\vec R=\{\vec r_1...\vec r_n\}$ are the coordinates of the system,
and the propagation in imaginary time $\tau$ is obtained by solving
\begin{equation}
\Psi(\vec R,\tau)=\int G(\vec R,\vec R',\tau)\Psi(\vec R',0)d\vec R' \,.
\end{equation}
The many-body Green's Function $G(\vec R,\vec R',\delta\tau)$ is accurately
known in the limit of small $\delta\tau$, and then the above integration 
can be solved using many small time steps. 
For spin-isospin independent interactions it is easy to write
\begin{equation}
e^{-H\delta\tau}\sim e^{-V(\vec R)\delta\tau}G_0(\vec R,\vec R',\delta\tau) \,,
\end{equation}
where $G_0$ is the free particle propagator that can be easily sampled as a 
particle diffusion step. The central potential is diagonal in the spin and
coordinates of particles, and can be interpreted as a probability factor.

For spin-isospin dependent or non-local potentials, the factor 
$\exp(-V(\vec R)\delta\tau)$ cannot be interpreted as a simple weight factor, but needs extra care.
In the GFMC all the spin-isospin states of the nucleons are explicitly included
in the wave function, and the potential has the effect of changing the 
amplitudes of each state~\cite{Carlson:1987}. The idea of AFDMC is instead to rewrite the
spin-dependent part of the potential as
\begin{equation}
\label{eq:prop}
V=\sum_{i<j}\sigma_{i\alpha}A_{i\alpha,j\beta}\sigma_{j\beta}=
\frac{1}{2}\sum_{n=1}^{3N}\lambda_n O_n^2 \,,
\end{equation}
where $\lambda_n$ are eigenvectors of the matrix $A_{i\alpha,j\beta}$,
and the $O_n$ operators are linear combinations of eigenvectors $\psi_n$ and 
spin operators and acting over particles:
\begin{equation}
O_n = \sum_i\sigma_{i\alpha}\psi_n^{i\alpha} \,.
\end{equation}

Using Eq.(\ref{eq:prop}) it is possible to linearize the spin operators 
in the propagator $\exp(-V(\vec R)\delta\tau)$ using the Hubbard-Stratonivich
transformation
\begin{equation}
e^{-\frac{1}{2}\lambda_n O_n^2\delta \tau} = 
\frac{1}{\sqrt{2\pi}}\displaystyle\int_{-\infty}^{+\infty}dx_n\;
e^{-\frac{x_n^2}{2}
-\sqrt{-\lambda_n\delta\tau} x_n O_n} \,.
\label{HS}
\end{equation}
The new variables $x_n$ are called auxiliary fields, and can be sampled 
using a Gaussian distribution.
The effect of the $O_n$ (linear) operators is to rotate the
single-particle spinors, and a more efficient wave function can be used.

Note that a three-neutron force can be rewritten in a quadratic
spin-operators form, and then its inclusion in the AFDMC propagator
is straightforward without any approximation. For more details see for
example Ref.~\cite{Gandolfi:2009} and references therein.

The Green's Function Monte Carlo (GFMC) is very accurate in the 
study of properties of light nuclei. The variational wave function
includes all the possible spin/isospin states of nucleons and it
provides a good variational ansatz to start the projection in the
imaginary time. The exponential growing of this states limits the present
calculations to $A \leq 12$ nuclei~\cite{Pieper:2005} or up to 16 neutrons~\cite{Gandolfi:2011}.
Using AFDMC the calculation can be extended up to $\sim 100$ neutrons,
making the simulation of homogeneous matter possible.
Although the implementation of GFMC and AFDMC are quite different,
they agree very well in computing the energy of 
neutrons in an external potential~\cite{Gandolfi:2011}, and
for 14 neutrons in a periodic box~\cite{Gandolfi:2009}.

\section{The equation of state of pure neutron matter}

We performed simulation of 66 particles in a box imposing periodic
boundary conditions. This number of fermions is large enough to guarantee
the thermodynamic limit. Finite size effects might have two contributions,
from the kinetic and the potential energy, whose effects have been 
carefully analyzed in previous work~\cite{Sarsa:2003,Gandolfi:2009}. The potential
energy is calculated by summing over periodic boxes as discussed in
Ref.~\cite{Sarsa:2003}. In any case the nuclear potential is very
short-range, and becomes almost zero well within the simulation box.
The kinetic energy size effects are kept under control by simulating
the system with 66 fermions. The energy of 66 non-interacting fermions
is very similar to the energy of the infinite system. In
Ref.~\cite{Gandolfi:2009} a careful analysis has been made using the
twist average boundary conditions, that guarantee a better extrapolation
to the thermodynamic limit, showing that the system is large enough.
In addition, in Refs.~\cite{Forbes:2011,Carlson:2011} it has been shown
that for the unitary Fermi gas, a simulation with more than 40 particles
essentially describes the infinite system.

\begin{table}
\begin{center}
\begin{tabular}{c|ccc}
\hline
$\rho$ [fm$^{-3}$] & \avep      & \avepuix  \\
\hline
0.04               &  6.55(1)  &  6.79(1) \\
0.05               &  7.36(1)  &  7.73(1) \\
0.06               &  8.11(1)  &  8.65(1) \\
0.07               &  8.80(1)  &  9.57(1) \\
0.08               &  9.47(1)  & 10.49(1) \\
0.09               & 10.12(1)  & 11.40(1) \\
0.10               & 10.75(1)  & 12.39(1) \\
0.11               & 11.37(1)  & 13.39(1) \\
0.12               & 12.00(1)  & 14.42(1) \\
0.13               & 12.64(1)  & 15.52(1) \\
0.14               & 13.21(1)  & 16.66(1) \\
0.15               & 13.84(2)  & 17.87(2) \\
0.16               & 14.47(2)  & 19.10(2) \\
0.20               & 17.11(2)  & 24.83(3) \\
0.24               & 19.98(3)  & 31.85(3) \\
0.28               & 23.00(3)  & 40.09(4) \\
0.32               & 26.45(3)  & 49.86(5) \\
0.40               & 34.06(5)  & 74.19(5) \\
0.48               & 42.99(8)  &105.9(1)  \\
\hline
\end{tabular}
\caption{Equation of state of neutron matter as a function of the density
for various Hamiltonians; statistical errors are shown in parentheses.
}
\label{tab:eos}
\end{center}
\end{table} 

\begin{figure}
\begin{center}
\includegraphics[width=0.9\columnwidth]{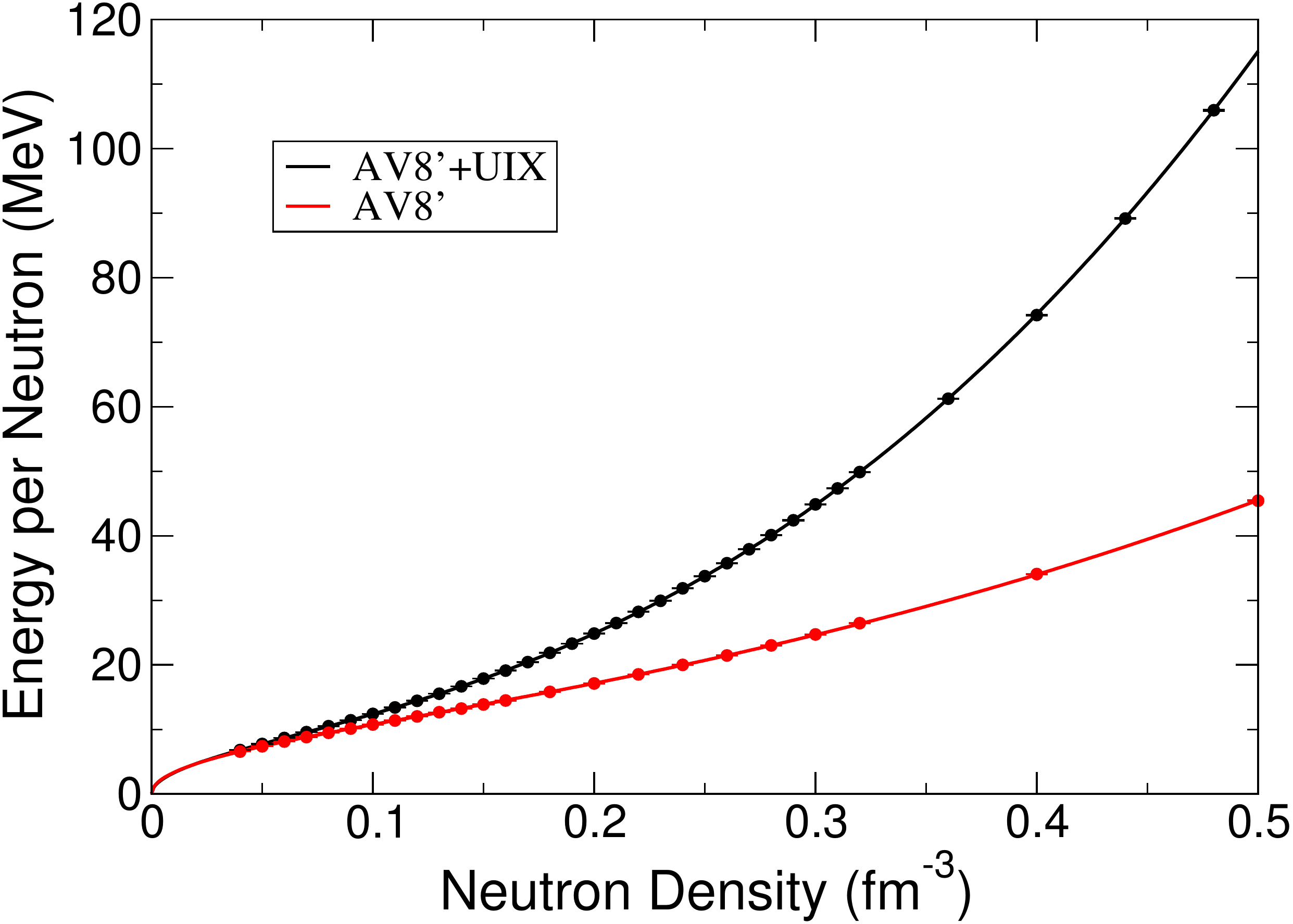}
\end{center}
\caption{The equation of state of neutron matter obtained 
using the \avep (two-body force only) and the \avepuix Hamiltonians.
}
\label{fig:eos1}
\end{figure}

The EoS of neutron matter obtained using the \avep
two-body force alone, and combined with the UIX three-body force, is presented
in Table~\ref{tab:eos}. The repulsion given by the addition of UIX is evident,
and becomes negligible at low densities. The results are also shown in
Fig.~\ref{fig:eos1}, and they are very similar to those obtained by Akmal, 
Pandharipande and Ravenhall using the AV18 potential~\cite{Akmal:1998}.

When the UIX three-body force is included in the calculation, we
find that the contribution of the Fujita-Miyazawa term
is very small in pure neutron matter,
while the short-range term is the dominant one. Similar contributions
of the two parts of UIX have also been calculated in Ref.~\cite{Akmal:1998}.
If instead, we calculate
the expectation value of UIX using a Fermi gas wave function without
correlations generated by the two- and three-body forces, we find a larger
contribution of the two-pion exchange term, similar to the Hartree-Fock
calculation of Ref.~\cite{Hebeler:2010}.

We find that the EoS can be accurately parametrized using the 
following functional form:
\begin{equation}
E(\rho_n)=a\left(\frac{\rho_n}{\rho_0}\right)^\alpha+
b\left(\frac{\rho_n}{\rho_0}\right)^\beta \,,
\label{eq:fit}
\end{equation}
where $E(\rho_n)$ is the energy per neutron as a function of the 
neutron density $\rho_n$, and the parameters $a$, $\alpha$, $b$, and $\beta$
are obtained by fitting the QMC results. 
The parametrization of the
equations of state obtained with the \avep and \avepuix Hamiltonians
are reported in Table~\ref{tab:fit}.

\begin{table*}[htbp]
\centering
\begin{tabular}{@{} lcccccccc @{}}
\hline
$3N$ force       & $E_{\rm sym}$ &$L$ & $a$    &  $\alpha$ & $b$ & $\beta$ & $\tilde a$ & $\tilde b$\\
               & (MeV)  & (MeV) & (MeV)    &        & (MeV)  &  & (MeV) & (MeV) \\
\hline
none                        & 30.5 & 31.3 & 12.7 & 0.49  & 1.78 & 2.26 & -27.4 & 6.2 \\
$V_{2\pi}^{PW}+V^R_{\mu=150}$&32.1 & 40.8 & 12.7 & 0.48  & 3.45 & 2.12 & -28.0 & 8.8 \\
$V_{2\pi}^{PW}+V^R_{\mu=300}$&32.0 & 40.6 & 12.8 & 0.488 & 3.19 & 2.20 & -28.2 & 8.9 \\
$V_{3\pi}+V_R$              & 32.0 & 44.0 & 13.0 & 0.49  & 3.21 & 2.47 & -29.2 & 9.9 \\
$V_{2\pi}^{PW}+V^R_{\mu=150}$&33.7 & 51.5 & 12.6 & 0.475 & 5.16 & 2.12 & -28.3 & 10.7 \\
$V_{3\pi}+V_R$              & 33.8 & 56.2 & 13.0 & 0.50  & 4.71 & 2.49 & -29.7 & 12.2 \\
UIX                         & 35.1 & 63.6 & 13.4 & 0.514 & 5.62 & 2.436& -29.8 & 13.6 \\
\hline
\end{tabular}
\caption{Fitting parameters for the neutron matter EoS defined in Eqs.(\ref{eq:fit}) and
(\ref{eq:fit2}) for selected different Hamiltonians.}
\label{tab:fit}
\end{table*}

In order to address the role and the uncertainties due to the
three-body force, very different models have been considered in
Ref.~\cite{Gandolfi:2012}. The main contribution of the three-body force is given
by the short-range part, whose structure in the UIX model is
\begin{equation}
\label{eq:vr}
V_{ijk}^R=A_R \sum_{\rm cyc} T^2(m_\pi r_{ij})T^2(m_\pi r_{jk}) \,,
\end{equation}
where $m_\pi$ is the pion mass, and
\begin{equation}
T(x)=\left(1+\frac{3}{x}+\frac{3}{x^2}\right)\frac{e^{-x}}{x}\xi^2(x) \,.
\label{eq:tx}
\end{equation}
The function $\xi(x)=1-e^{-cx^2}$ is a cutoff function to regularize
$T(x)$ at small distances. In order to address the role of the above 
short-distance behavior, we have also considered very different forms to
replace $T(x)$ in Eq.(\ref{eq:vr}):
\begin{eqnarray}
f_1(x)&=&\exp(-2\mu x) \ , \nonumber \\
f_2(x)&=&\frac{\exp(-2\mu x)}{\mu x} \left[1-\exp(-(2\mu x)^2)\right] \ ,
\end{eqnarray}
where the $\mu$ parameter to control the effective range of the short-range
repulsion has been varied from 150 to 600 MeV. In addition we have also
replaced the general form of Eq.(\ref{eq:vr}) with
\begin{equation}
V_{ijk}^R=A_R f(r_{ij})f(r_{jk})f(r_{ki}) \,,
\label{eq:vrprod}
\end{equation}
to explore the sensitivity to the geometry of the three-neutron configuration.

For each different function $f(r)$ and different $\mu$, we have fixed
the strength $A_R$ to reproduce a particular energy of neutron matter
at saturation density $\rho_0=0.16$ fm$^{-3}$. In this way we can relate
the EoS to the symmetry energy that we shall describe in the next section.

In addition to the effects due to the short-range part,
we have also explored the effect of intermediate- and long-range
contributions of the three-body force. These terms, where two or
three pions are exchanged between neutrons, with the creation of
$\Delta$ excited states, are described in Ref.~\cite{Pieper:2001}. We
have considered different combinations of these terms with the above
short-range parts in order to explore various models of three-body force
with very different ranges.
Also different values of the cutoff $c$ entering in Eq.(\ref{eq:tx}), and 
in the two- and three-pion exchange have been considered. 

\begin{figure}
\begin{center}
\includegraphics[width=0.9\columnwidth]{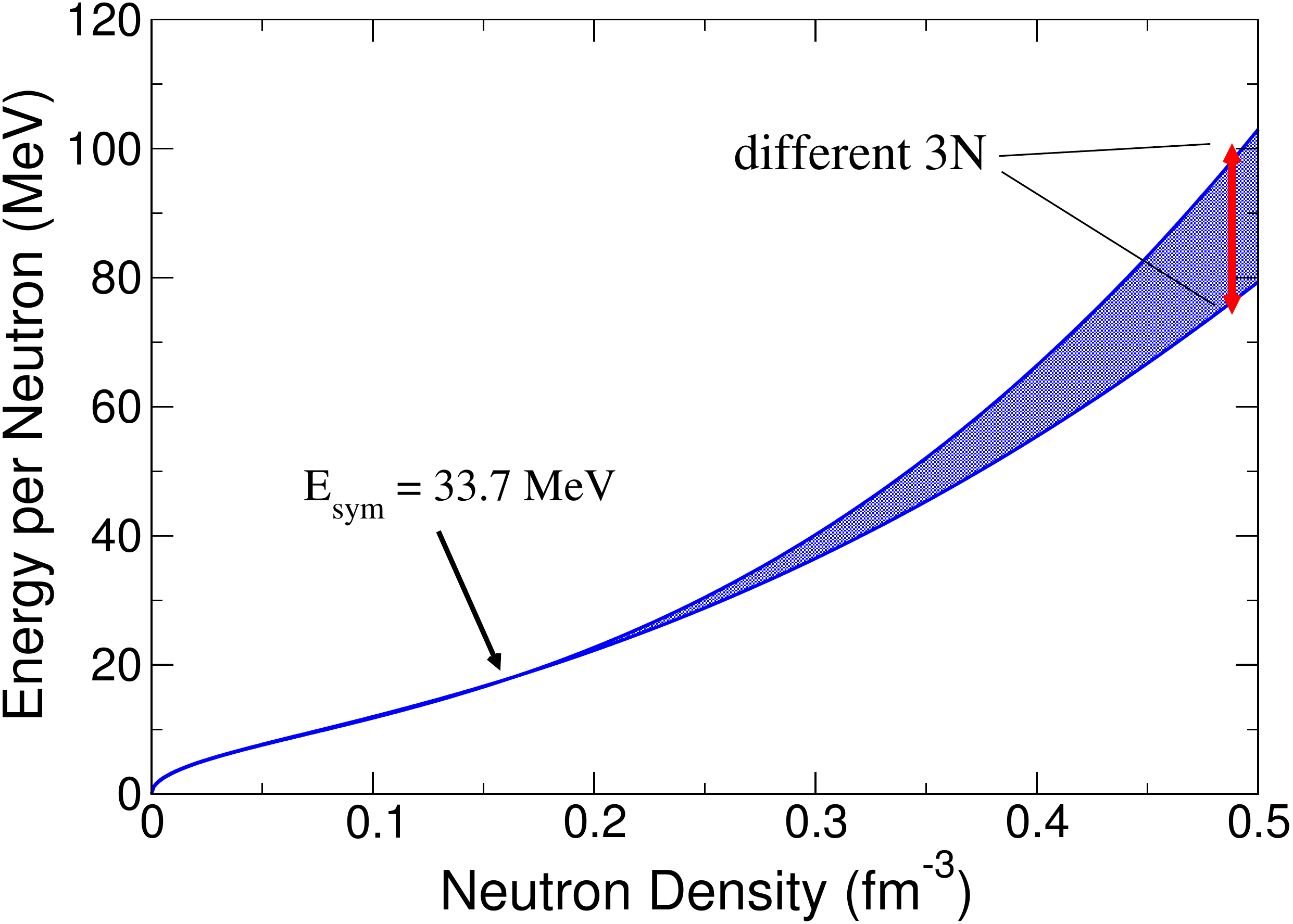}
\end{center}
\caption{The equation of state of neutron matter obtained by
using various models of three-neutron force as described in the 
text. For each model we impose that the energy at saturation is
17.7(1) MeV.}
\label{fig:eos2}
\end{figure}

All the various EoS obtained with the different
three-neutron forces are grouped in the band showed in
Fig.~\ref{fig:eos2}, and they give the same energy at saturation
corresponding to 17.7(1) MeV.  The various nuclear Hamiltonians
give EoS with a very similar behavior up to about 0.24 fm$^{-3}$, 
and for higher densities the result becomes model-dependent. 
Within our model this analysis permits us to understand the
uncertainties associated with the model of the three-neutron force, but 
it will be important to explore the effect of other diagrams predicted
by $\chi$EFT not included in the Illinois models.

\begin{figure}
\begin{center}
\includegraphics[width=0.9\columnwidth]{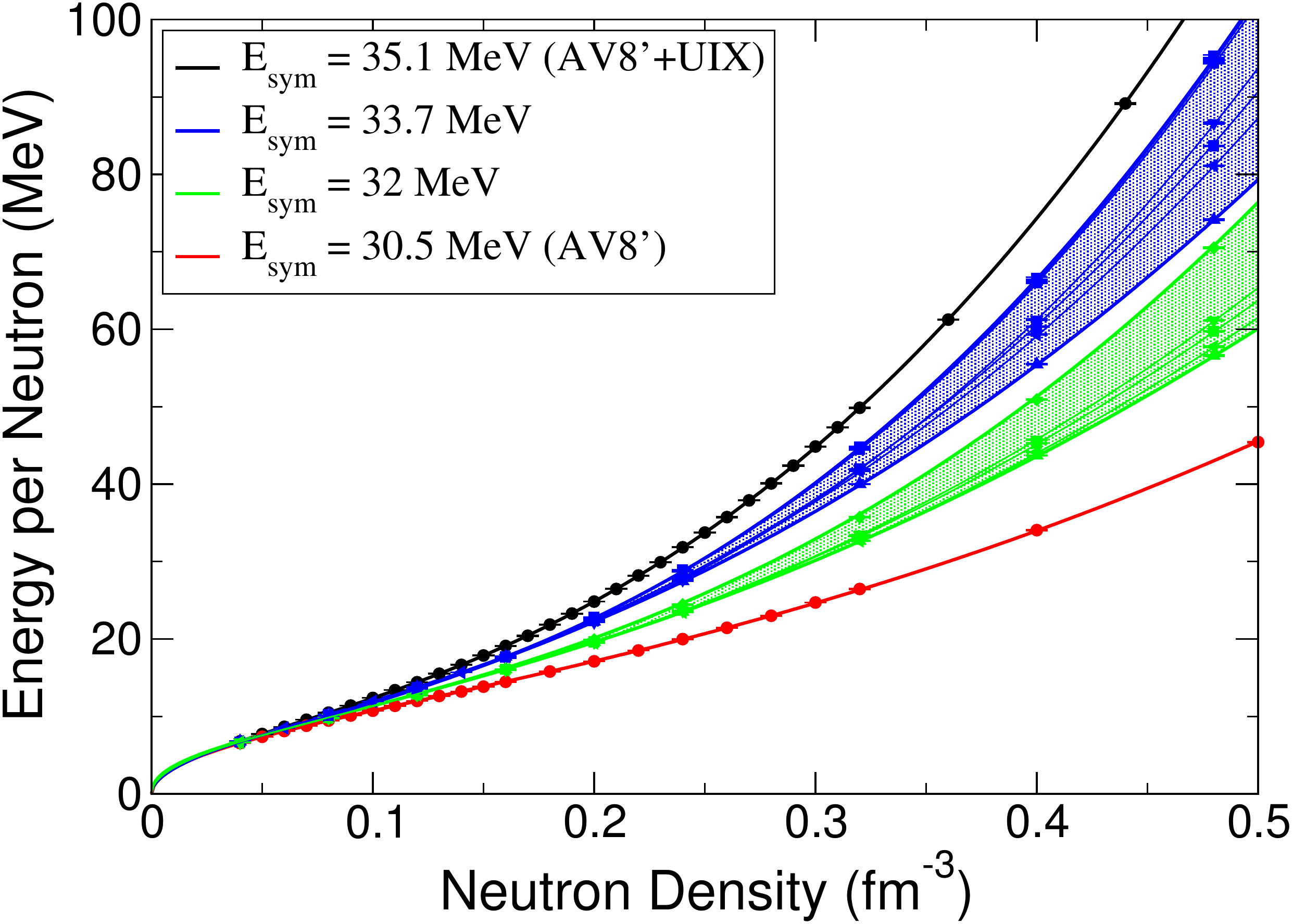}
\end{center}
\caption{The equation of state of neutron matter obtained by
using various models of three-neutron force as described in the 
text. For each model we impose that the energy at saturation is
17.7(1) MeV (blue band), or 16.0(1) MeV (green band). The results
are compared with the equations of state obtained with the \avep and
\avepuix Hamiltonians. In the legend we indicate the corresponding
symmetry energy at saturation. }
\label{fig:eosall}
\end{figure}

As an additional investigation, all the above models of three-neutron
force have also been adjusted to reproduce an energy of 16 MeV of
pure neutron matter. The results are shown in Fig.~\ref{fig:eosall},
where we compare with the results obtained with the \avep
and \avepuix Hamiltonians. The EoS given by \avep and \avepuix are shown
without error bars because they correspond to the original Hamiltonians
that we have considered.  They may be considered as extreme cases because they correspond
to an interval of the symmetry energy between 30.5 and 35.1 MeV, 
a range compatible with several experimental measurements~\cite{Tsang:2012,Lattimer:2012}.
(Error bands could be constructed by introducing a range
of three-nucleon forces with these symmetry energies, but we expect
the size of these error bands to be similar to the others.)
To obtain a lower symmetry energy with the \avep model of two-body force
would require an attractive contribution of the three-body force.
The three-pion rings give attraction in pure neutron matter~\cite{Maris:2013}, but they
would require a strong short-range repulsion to make the EoS stiff enough
to support astrophysical observations.

\section{Symmetry energy}

At saturation density, the EoS of pure neutron matter is useful to
extract information on the symmetry energy and its slope. The symmetry
energy, $E_{\rm sym}(\rho)$ is may be defined as the difference
between the energy per baryon of pure neutron matter and the energy
per baryon of infinite homogeneous nuclear matter with equal neutron
number density, $\rho_n$, and proton number density, $\rho_p$.
In terms of the isospin asymmetry,
$x\equiv\rho_p/\rho$, the energy per baryon of isospin asymmetric
nuclear matter can be expanded in even powers of $x$,
\begin{equation}
E(\rho,x)=E_0(\rho)+E_{\rm sym}^{(2)}(\rho)(1-2x)^2+
E_{\rm sym}^{(4)}(1-2x)^4+\dots \,,
\end{equation}
where $E$ is the energy per baryon of the system,
$E_0(\rho)=E(\rho,x=0.5)$ is the EoS of symmetric
nuclear matter, and $E_{\rm sym}^{(4)}(\rho)$ and higher order
corrections will be ignored. The symmetry energy $E_{\rm sym}$ is thus
given by
\begin{equation}
E_{\rm sym}(\rho)=E(\rho,0)-E_0(\rho) \,,
\end{equation}
where $E(\rho,0)$ is the EoS of pure neutron matter.
Near the nuclear saturation density, $\rho_0$, there are a number of
constraints on the EoS of infinite nuclear matter from
nuclear masses, charge radii, and giant resonances.
The extrapolation of the
binding energy of heavy nuclei to the thermodynamic limit yields
$E_0(\rho_0)=-16.0\pm0.1$ MeV~\cite{Moller:1995}. Because the pressure is zero at saturation, we
can expand the symmetry energy around saturation $\rho=\rho_0$ as
\begin{equation}
E_{\rm sym}(\rho)\equiv E_{\rm sym}+
\frac{L}{3}\frac{\rho-\rho_0}{\rho_0}+\dots \,,
\label{eq:lvsesym}
\end{equation}
where $E_{\rm sym}$ is the symmetry energy at saturation
(sometimes called $S$), and $L$ is a parameter related to the 
slope of $E_{\rm sym}(\rho)$.

\begin{figure}
\begin{center}
\includegraphics[width=0.9\columnwidth]{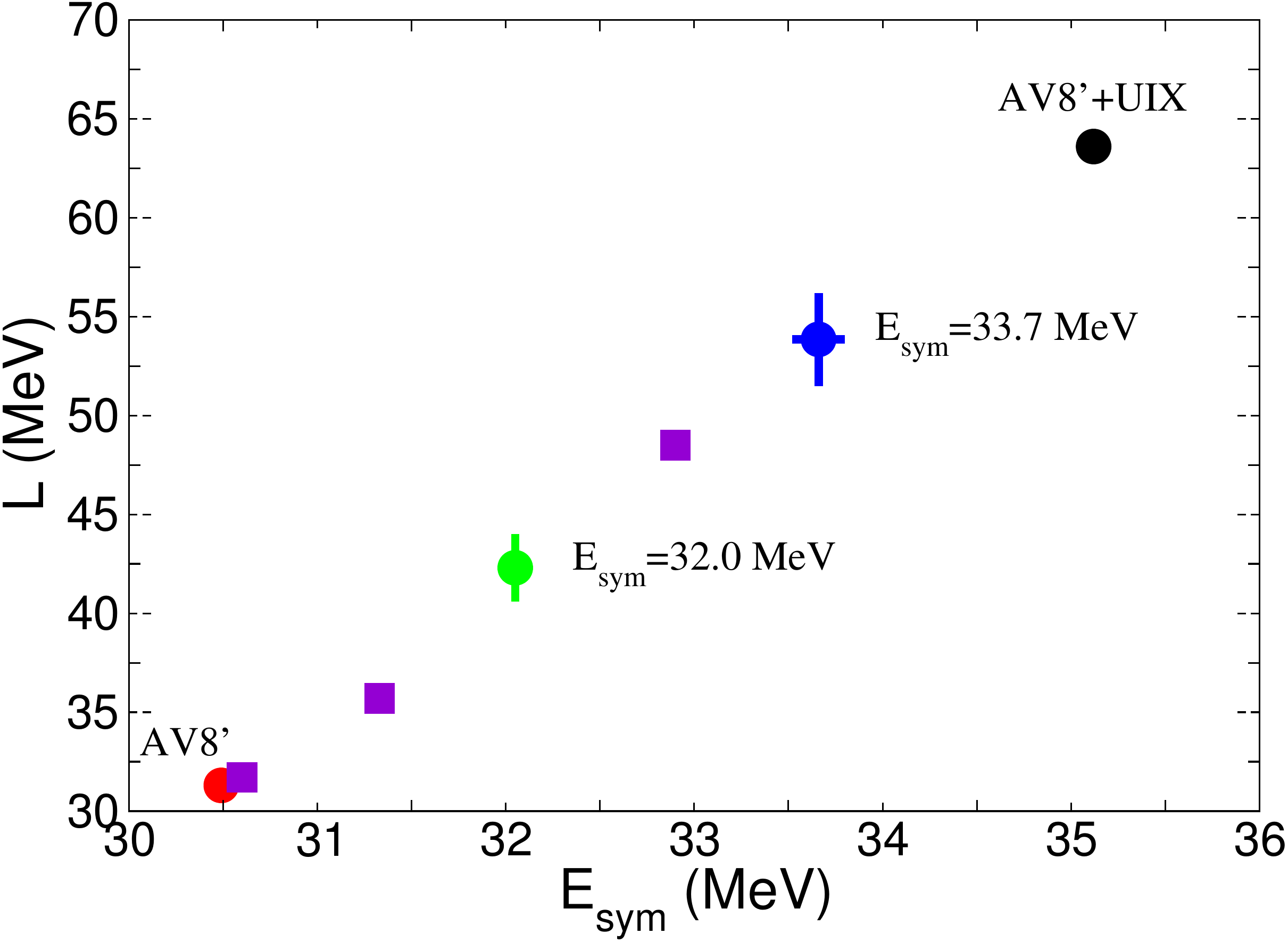}
\end{center}
\caption{The value of L as a function of E$_{\rm sym}$ obtained from
various EoS. The green and blue points with error bars correspond to
the various EoS indicated by the two colored areas of
Fig.~\ref{fig:eosall}. The red and black points show the results
based on the EoS presented in Fig.~\ref{fig:eos1}, i.e. the results
obtained using a two-body force alone and combined with the UIX
model. The squared symbols correspond to results obtained by
independently changing the cutoff parameters entering in $V_R$ and
in the three-pion rings of the three-neutron force.
}
\label{fig:lvsesym}
\end{figure}

Using Eq.(\ref{eq:lvsesym}) we can fit the value of $E_{\rm sym}$ and L
to the calculated EoS as described in the previous section. The result
is summarized in Fig.~\ref{fig:lvsesym}, where we compare the results
obtained using the \avep and \avepuix Hamiltonians (red and black
symbols), the various EoS giving the indicated $E_{\rm sym}$ obtained by
changing the three-neutron force model (green and blue symbols), and
results obtained using the Illinois model of three-neutron force that
includes three-pion rings where we have independently changed the
cutoff of the intermediate- and short-range part. It is clear that
within our model the correlation between L and $E_{\rm sym}$ is very
strong, and following a linear trend. Future work will need to
investigate the role of the two-body force that in this work has not
been changed.

From the functional form in Eq.(\ref{eq:fit}), the symmetry
energy and its slope at saturation $\rho_0$ are given by
\begin{equation}
E_{\rm sym}=a+b+16 \,,
\end{equation}
and
\begin{equation}
L=3\,(a\alpha+b\beta) \,.
\end{equation}
The parameters of Table~\ref{tab:fit} clearly show that the parameters
$a$ and $\alpha$ are marginally dependent to the three-body force, and
thus a precise measurement of $E_{\rm sym}$ would in principle
constrain the strength of three-nucleon interactions.

An alternative to the parametrization of Eq.(\ref{eq:fit}) for the QMC results
can be obtained by separating the energy of the noninteracting Fermi gas
from the interaction in the following way:
\begin{equation}
E(\rho_n)=E_{FG}+
\tilde a\left(\frac{\rho_n}{\rho_0}\right)+
\tilde b\left(\frac{\rho_n}{\rho_0}\right)^2 \,,
\label{eq:fit2}
\end{equation}
where $E_{FG}=\frac{3}{5}\frac{\hbar^2}{2m}\left(3\pi^2\rho_n\right)^{2/3}$.
The parameters obtained by fitting the QMC results using the above functional
form are also reported in Table~\ref{tab:fit}.
Using this expression, the quality of the fit is slightly worse than using
the expression of Eq.(\ref{eq:fit}), but in this case we have
only two free parameters instead of four.
The symmetry energy and its slope obtained from this alternative expression 
can be obtained as
\begin{equation}
E_{\rm sym}=E_{FG}^0+\tilde a+\tilde b+16 \,,
\end{equation}
and
\begin{equation}
L=3\,\left(\frac{2}{3}E_{FG}^0+\tilde a+2\,\tilde b\right) \,,
\end{equation}
where $E_{FG}^0$ is the Fermi gas energy at density 0.16 fm$^{-3}$.
Using the values reported in Table~\ref{tab:fit} both $E_{\rm sym}$ and L
are very close to the values reported in the same table and in Fig.~\ref{fig:lvsesym}. 

\section{Neutron star masses, radii and the EoS}
\label{sec:GCR}

The EoS of neutron matter is the principal ingredient
to study the structure of neutron stars. The neutron star matter is 
mainly composed by neutrons and a few protons. Once the EoS
is given, the mass-radius (M-R) relation of neutron stars can be obtained by
by integrating the Tolman-Oppenheimer-Volkoff (TOV) equations:
\begin{equation}
\frac{dP}{dr}=-\frac{G[m(r)+4\pi r^3P/c^2][\epsilon+P/c^2]}{r[r-2Gm(r)/c^2]} \,,
\label{eq:tov1}
\end{equation}
\begin{equation}
\frac{dm(r)}{dr}=4\pi\epsilon r^2 \,,
\label{eq:tov2}
\end{equation}
where $P=\rho^2(\partial E/\partial\rho)$ and $\epsilon=\rho(E+m_N)$
are the pressure and the energy density, $m_N$ is the neutron mass,
$m(r)$ is the gravitational mass enclosed within a radius $r$, and $G$
is the gravitational constant. The solution of the TOV equations for a
given central density gives the profiles of $\rho$, $\epsilon$ and $P$
as functions of radius $r$, and also the total radius $R$ and mass
$M=m(R)$. The total radius $R$ is given by the condition $P(R)=0$.
The TOV equations are modified only slightly by magnetic fields
and temperatures which are expected, and rotation is less than a
10\% effect for the kinds of M-R curves which we obtain below.
Thus to first order, all neutron stars are expected to lie on one
M-R curve determined entirely by the EoS of cold dense matter.
Also, an ensemble of neutron star mass and radius measurements
which determine the M-R curve constrain the EoS. The speed of
sound, $c_s$ in the neutron star interior is $c_s^2 =
dP/d\epsilon$, and ensuring that this is less than the speed of
light (and thus the EoS is said to be ``causal'') constrains the
set of possible EoS. Also, the pressure must increase with
increasing energy density, $dP/d\epsilon>0$, in order to ensure
that the neutron star is hydrodynamically stable.

The neutron star mass measurements which provide the strongest
EoS constraints are those which have the highest mass. Recent
observations~\cite{Demorest:2010,Antoniadis13} have found
two neutron stars almost 2 $M_{\odot}$. These two data points
provide some of the strongest constraints on the nature of
zero-temperature QCD above the nuclear saturation density.

\begin{figure}
\begin{center}
\includegraphics[width=0.95\columnwidth]{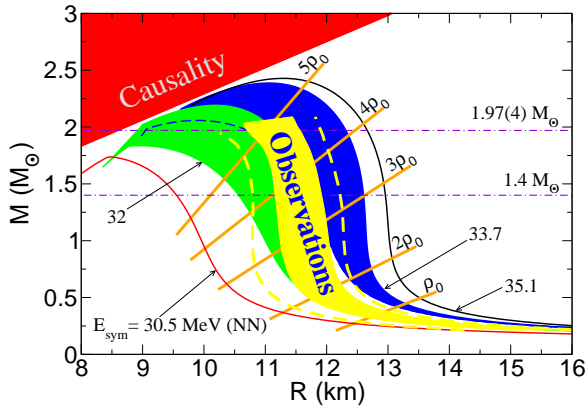}
\end{center}
\caption{The mass-radius relation for neutron stars based on the QMC
neutron matter results above. Results are presented for the
different EoS given in Fig.~\ref{fig:eosall}. The numbers indicate
the value of $E_{\rm sym}$ of the various EoS. The yellow band
corresponds to observations of neutron stars obtained in
Ref.~\cite{Steiner:2012}.}
\label{fig:mr}
\end{figure}

We begin by examining what can be deduced about the M-R relation
directly from the neutron matter EoS using the 1.97(4)
M$_\odot$~\cite{Demorest:2010} measurement, but without employing a
separate model for high-density matter. Using the form in
Eq.(\ref{eq:fit}), the energy density and pressure are given by
\begin{equation}
\epsilon=\rho_0\left[a\left(\frac{\rho}{\rho_0}\right)^{1+\alpha} 
+b\left(\frac{\rho}{\rho_0}\right)^{1+\beta}+m_n\left(\frac{\rho}
{\rho_0}\right)\right]\,,
\end{equation}
and
\begin{equation} 
P=\rho_0\left[a\alpha\left(\frac{\rho}{\rho_0}\right)^{1+\alpha}
+b\beta\left(\frac{\rho}{\rho_0}\right)^{1+\beta}\right]\,.
\end{equation}
We solve the TOV equations using QMC calculations shown in the above
sections for $\rho \ge \rho_{\rm crust}=0.08$ fm$^{-3}$. For lower
densities we use the EoS of the crust obtained in Refs.
\cite{Baym:1971} and \cite{Negele:1973}. At high densities we use the
maximally stiff EoS when the QMC models violate the causality and
become superluminal. The results of the M-R diagram of neutron stars
is presented in Fig.~\ref{fig:mr}, where we compare the results given
by the various EoS described above. The numbers in the figure
indicate the symmetry energy associated with the various equations of
state. In the figure we also indicate with the orange lines the
density of the neutron matter inside the star. Even at large masses
the radius of the neutron star is mainly governed by the equation of
state of neutron matter between 1 and 2 $\rho_0$~\cite{Lattimer:2001}.
These results are not qualitatively modified by corrections from a
non-zero proton fraction~\cite{Gandolfi:2010}.

The \avep Hamiltonian alone does not support the recent observed
neutron star with a mass of 1.97(4)M$_\odot$~\cite{Demorest:2010}.
However, adding a three-body force to \avep can provide sufficient
repulsion to be consistent with all of the
constraints~\cite{Gandolfi:2012}. There is a clear correlation
between neutron star radii and the symmetry energy which determines
the EoS of neutron matter between 1 and 2 $\rho_0$. The results in
Fig.~\ref{fig:mr} also show that the most modern neutron matter EoS
imply a maximum neutron star radius not larger than 13.5 km, unless
a drastic repulsion sets in just above the saturation density. This
rules out EoS with large values of $L$, typical of Walecka-type
mean-field models without higher-order meson couplings which can
decrease $L$.

\section{Radius measurements}
\label{sec:radius}

\begin{figure}
\begin{center}
\includegraphics[width=0.49\textwidth]{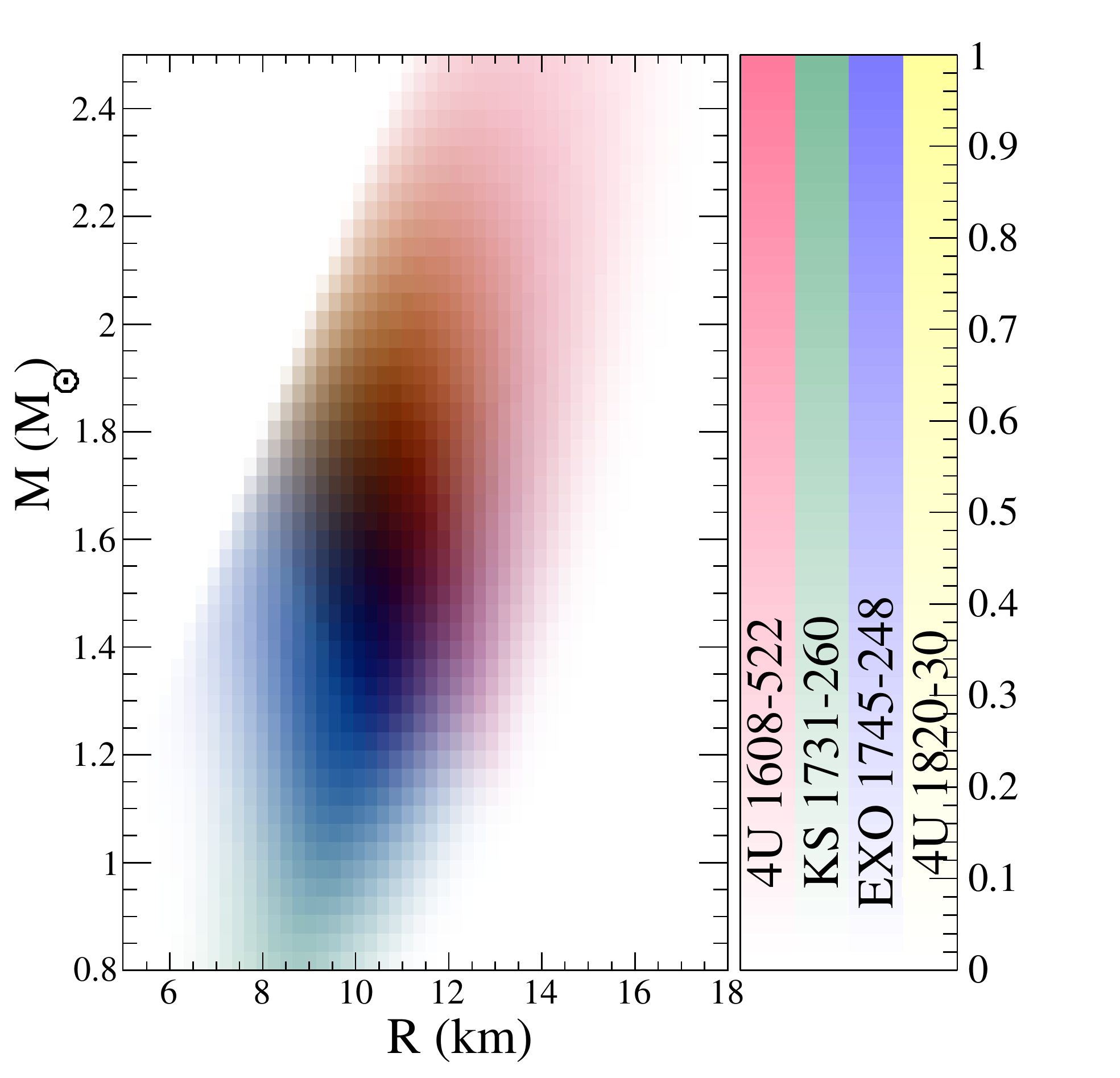}
\end{center}
\caption{Probability distributions in the mass-radius plane for four
  neutron stars exhibiting PRE X-ray bursts. Colors are added together
  in RGB color space.}
\label{fig:pre_ns}
\end{figure}

\begin{figure}
\begin{center}
\includegraphics[width=0.49\textwidth]{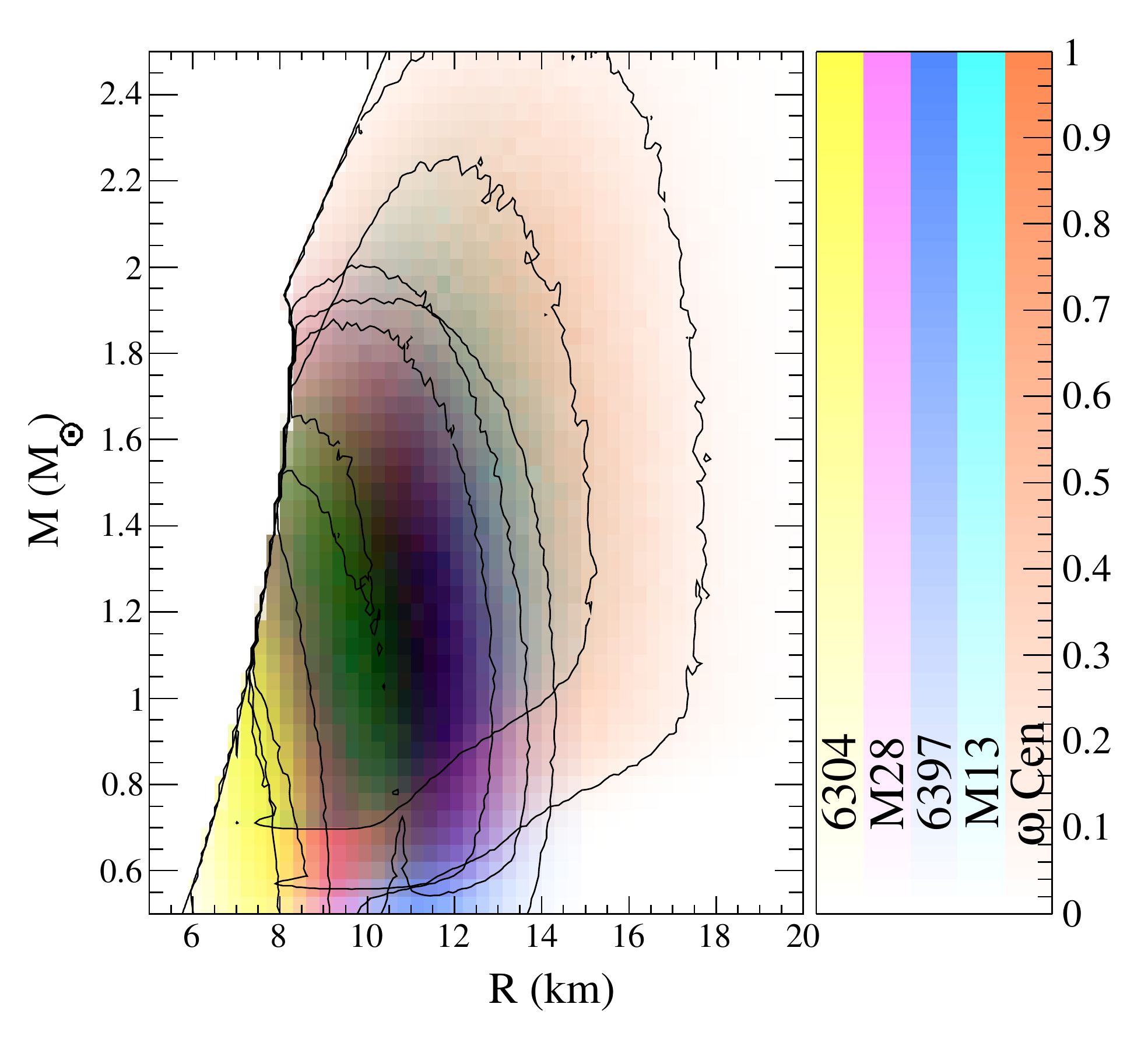}
\end{center}
\caption{Probability distributions in the mass-radius plane for five
  neutron stars in five globular clusters from Ref.~\cite{Lattimer13}.
  Colors are added together in RGB color space when necessary. The
  contour lines outline the 90\% confidence regions.}
\label{fig:qlmxb_ns}
\end{figure}

Neutron star radius measurements have proven more difficult, because
they require both a distance measurement and some degree of modeling
of the neutron star X-ray spectrum. Low-mass X-ray binaries (LMXBs)
are neutron stars accreting matter from a low mass main-sequence or
white dwarf companion. There are two types of LMXB observations which
have recently provided neutron star radius information. The first type
are LMXBs which exhibit photospheric radius expansion (PRE) X-ray
bursts, thermonuclear explosions strong enough to temporarily lift the
surface of the neutron star
outwards~\cite{vanParadijs:1979,Ozel:2010}. Several neutron stars have
exhibited PRE X-ray bursts and four which have have been used to infer
the neutron star mass and radius are given in Fig.~\ref{fig:pre_ns},
using the methods described in Ref.~\cite{Steiner:2010}. The second
type are quiescent LMXBs, (QLMXBs), where the accretion from the
companion has stopped, allowing observation of the neutron star
surface which has been heated by accretion~\cite{Rutledge:1999}. A
recent analysis of five neutron stars~\cite{Lattimer13} including the
possibility of both hydrogen and helium atmospheres and distance
uncertainties is shown in Fig.~\ref{fig:qlmxb_ns}. Note that
already from these two figures alone, it is clear that the 
data favors a radius near 11 km.

\section{Bayesian analysis}

In this section, we constrain the equation of state and symmetry
energy using observational data sets similar to that described in
Sec.~\ref{sec:radius} and include the possibility of phase
transitions in matter above the nuclear saturation density. In order
to do this, we parametrize the EoS of matter at higher densities with
a simple expression rich enough to include exotic matter. In our
fiducial model, we employ the parametrization in Eq.(\ref{eq:fit}) for
matter near the saturation density and treat the crust as before. At
some higher density $\rho_t \sim 0.24 - 0.48$ fm$^{-3}$ the EoS may
change due to the presence of exotic matter or a higher-order
many-body correction. Beginning with this density,
we employ a polytrope of the form $P=K_1 \epsilon^{\Gamma}$, fixing
$K_1$ to ensure that the EoS is continuous and setting
$\Gamma_1=1+1/n_1$ where $n_1$ is the ``polytropic index''. At a
higher energy density, $\epsilon_2$, we use a second polytrope with
index $n_2$, fixing $K_2$ to ensure that the EoS is continuous.

We perform a Bayesian analysis using data from QLMXBs and 
neutron stars which exhibit PRE bursts, where our model
space is given by the EoS parameters and also one parameter
for the mass of each neutron star in the data set. Given
an EoS, the TOV equations provide the M-R curve and thus
a prediction for the radius of each neutron star from
its mass. As described above, we always ensure that
our EoS are causal, hydrodynamically stable, and that
our M-R curves support a 1.97 solar mass star.
In addition to our fiducial model, we construct a EoS 
parametrization which describes a hybrid neutron star with
deconfined quark matter at the core. In this case, the 
higher-density polytrope is replaced by the quark
matter model of Ref.~\cite{Alford:2005}.

Finally, neutron stars contain a small amount of protons,
so we multiply the EoS by a small 
($\sim$ 10\%) and density-dependent correction factor
which modifies the pressure. This correction factor is
obtained by averaging over Skyrme forces which give similar
M-R curves to those suggested by the data.

\begin{figure}
\begin{center}
\includegraphics[width=0.9\columnwidth]{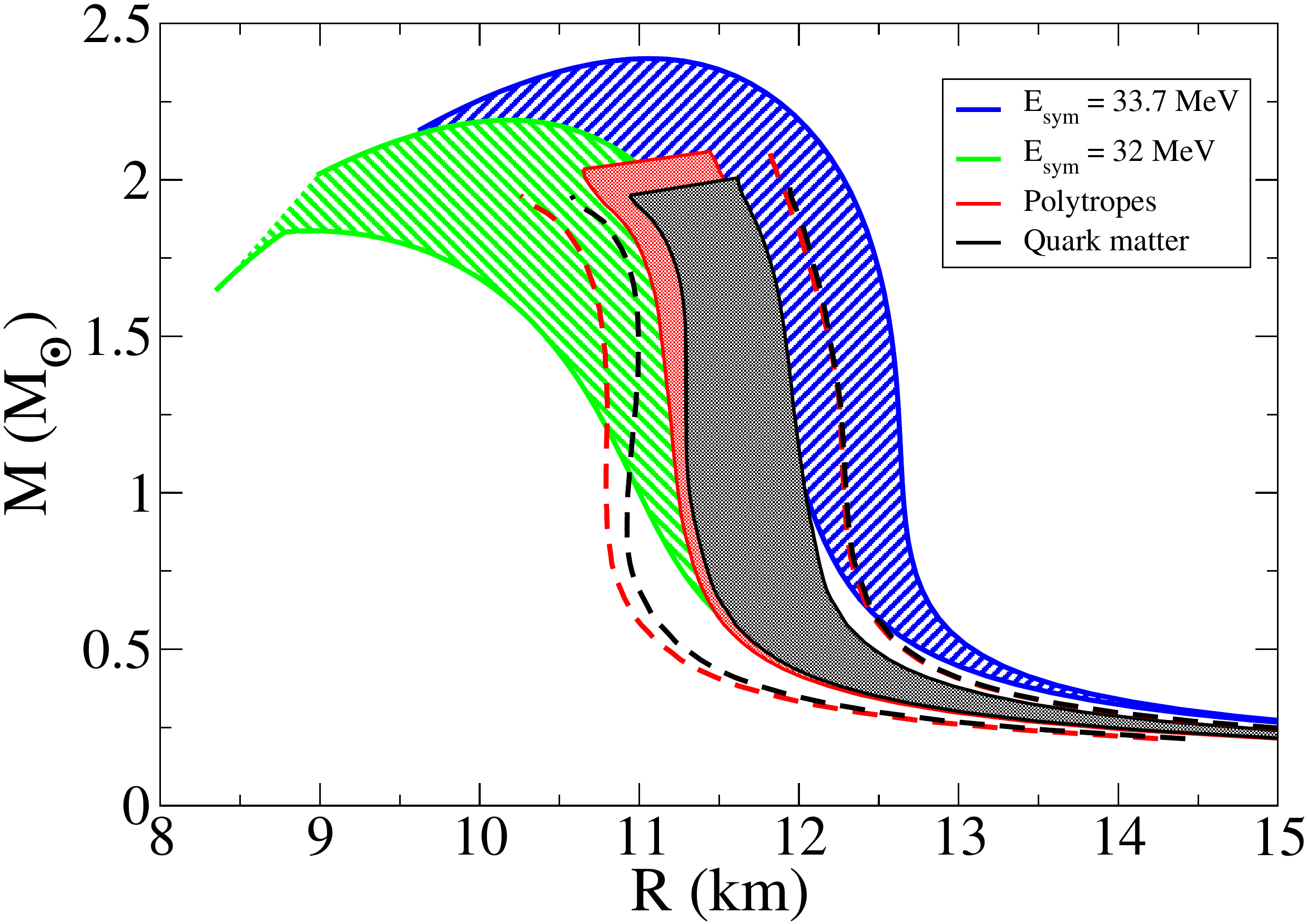}
\end{center}
\caption{The mass-radius relation obtained from the Bayesian analysis 
of data as described in the text. The red and black bands (dashed lines) 
are the results within 1$\sigma$ (2$\sigma$) obtained using different 
models for the parametrization of the high density EoS~\cite{Steiner:2012}. 
The green and blue bands are the results presented in Fig.~\ref{fig:mr}.
}
\label{fig:mr_comp}
\end{figure}

The main results are presented in Fig.~\ref{fig:mr_comp}, where
we show the M-R diagram obtained from eight neutron star data sets.
In the figure, the red and black bands correspond to the M-R profile
obtained within a 67\% of confidence, and dashed lines correspond to
98\%. The red and black bands are obtained by modeling the high
density part with two polytropes and with a polytrope and quark
matter, respectively. Generally, the additional constraint from the
observational neutron star data rules out larger radii, bringing the
maximum down to just over 12 km and giving a minimum down to about
10.5 km, independent of whether or not the neutron star contains
quark matter in the core. We also find that the effect of varying
$\rho_t$ between 0.24 and 0.48 fm$^{-3}$ is relatively small.

\begin{figure}
\parbox{1.65in}{
\vspace*{-1.15in}
\includegraphics[width=1.65in]{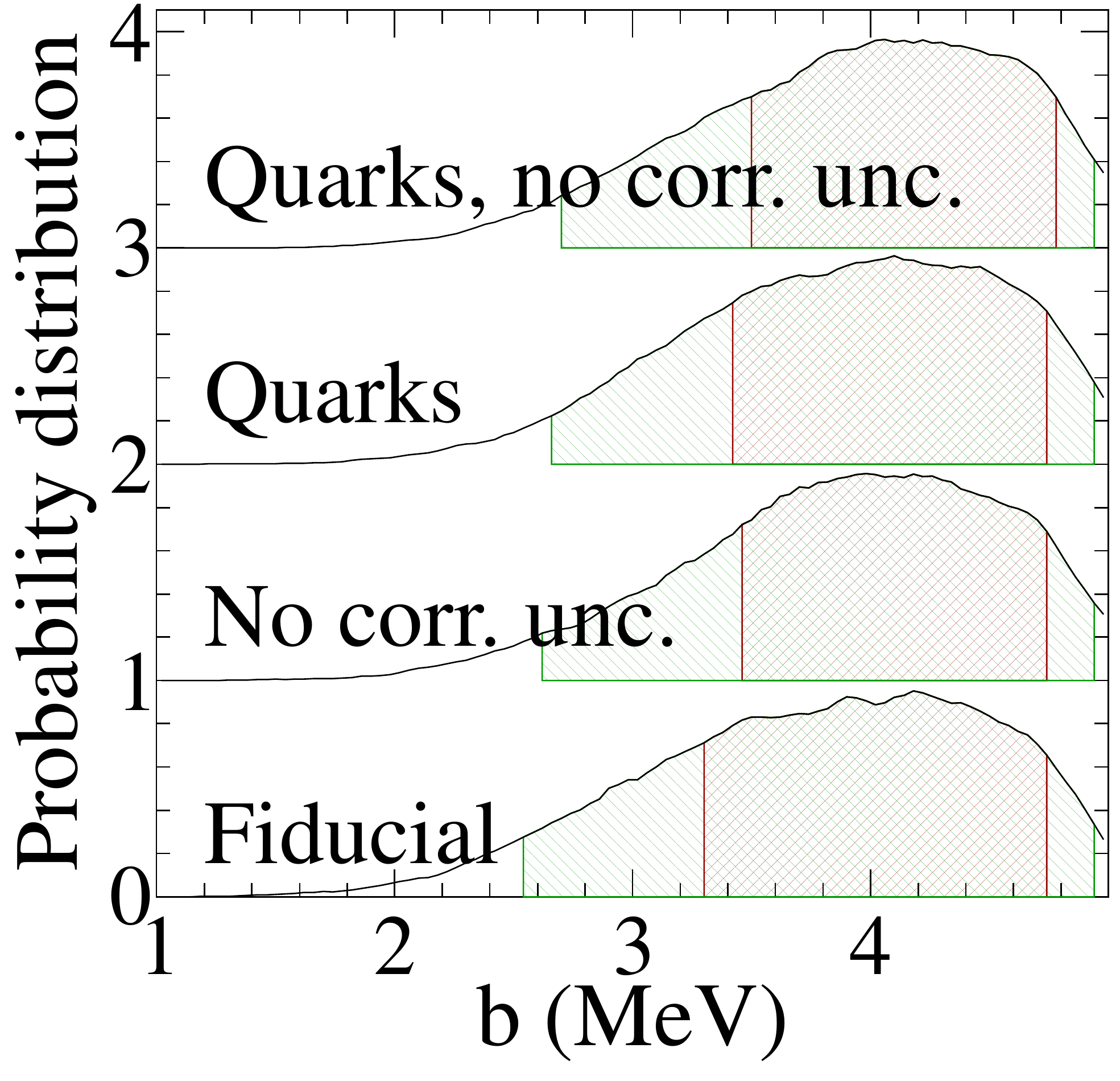}
\includegraphics[width=1.65in]{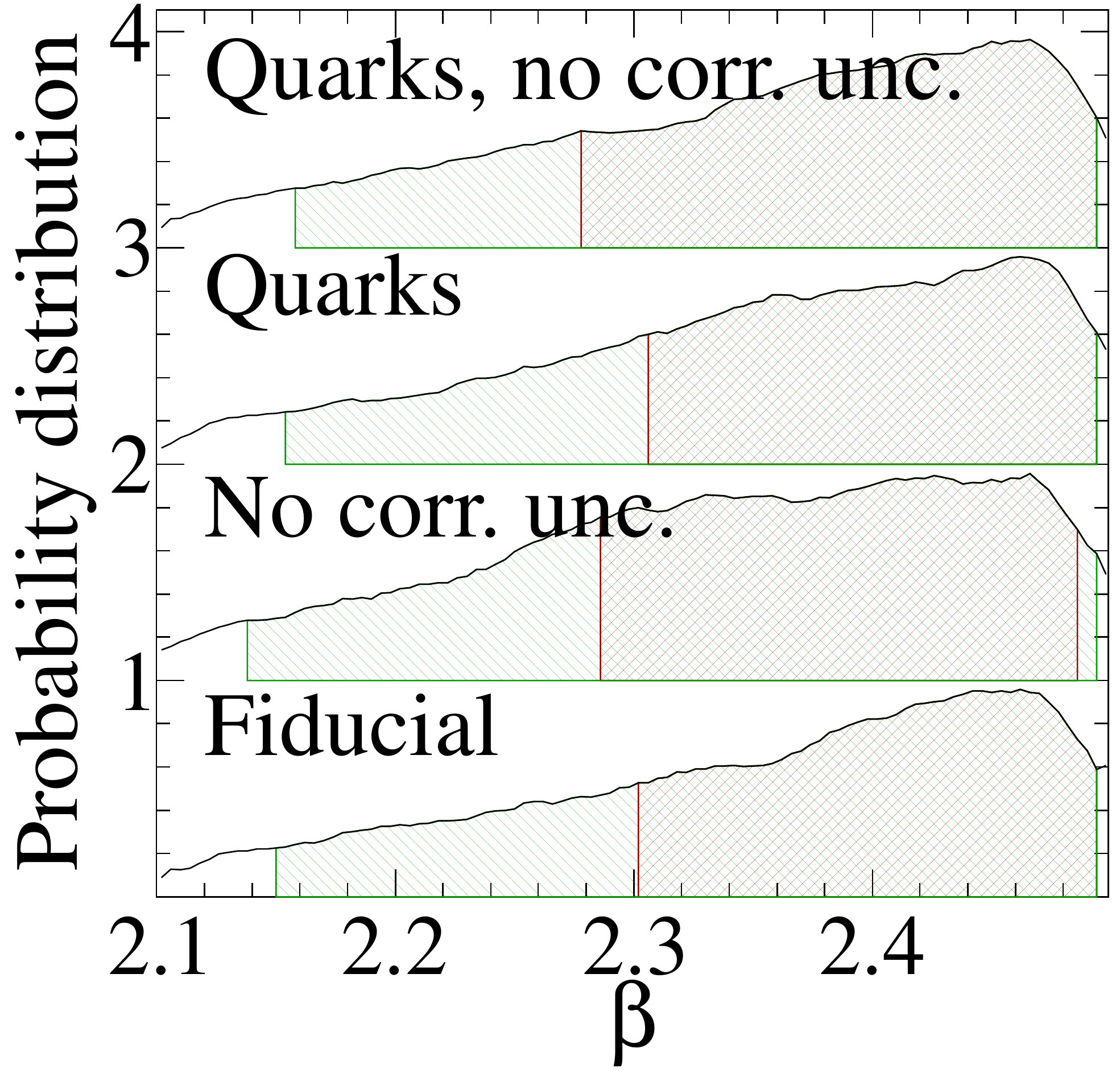}
}
\includegraphics[width=1.65in]{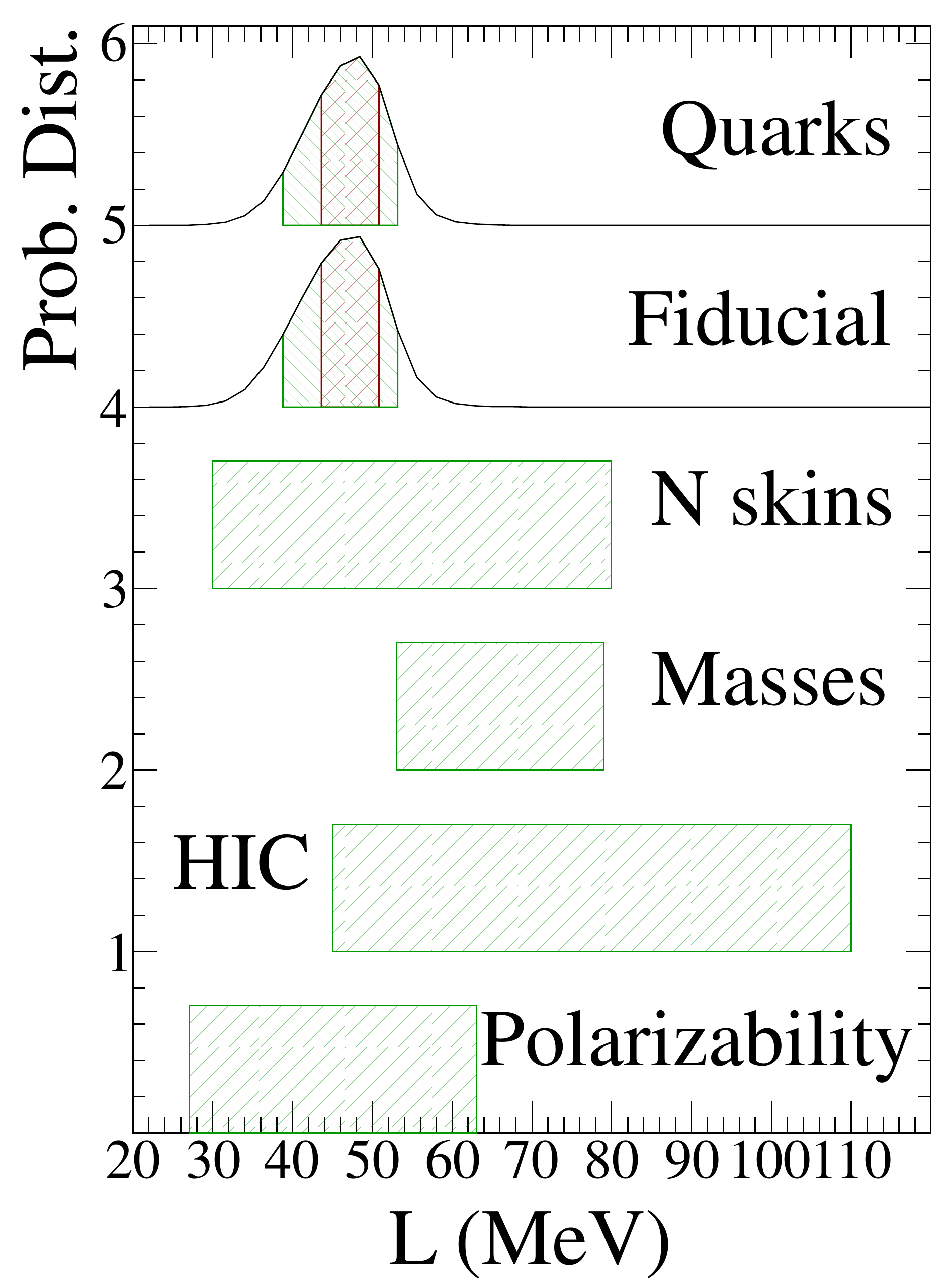}
\caption{The left panel shows probability distributions of the
  parameters $b$ and $\beta$ obtained from the Bayesian analysis. The
  right panel summarizes constraints on $L$ from observations and
  experiments. The top two curves show constraints on $L$ as
  probability distributions assuming either the fiducial model of
  Ref.~\cite{Steiner:2012} or the model containing quarks. The bottom
  four curves show constraints on $L$ from experiment, from neutron
  skins~\cite{Warda:2009}, nuclear masses~\cite{Liu:2010}, heavy-ion
  collisions~\cite{Tsang:2009}, and from the electric dipole
  polarizability~\cite{Tamii:2011}.}
\label{fig:params}
\end{figure}

\begin{figure}
\begin{center}
\includegraphics[width=0.9\columnwidth]{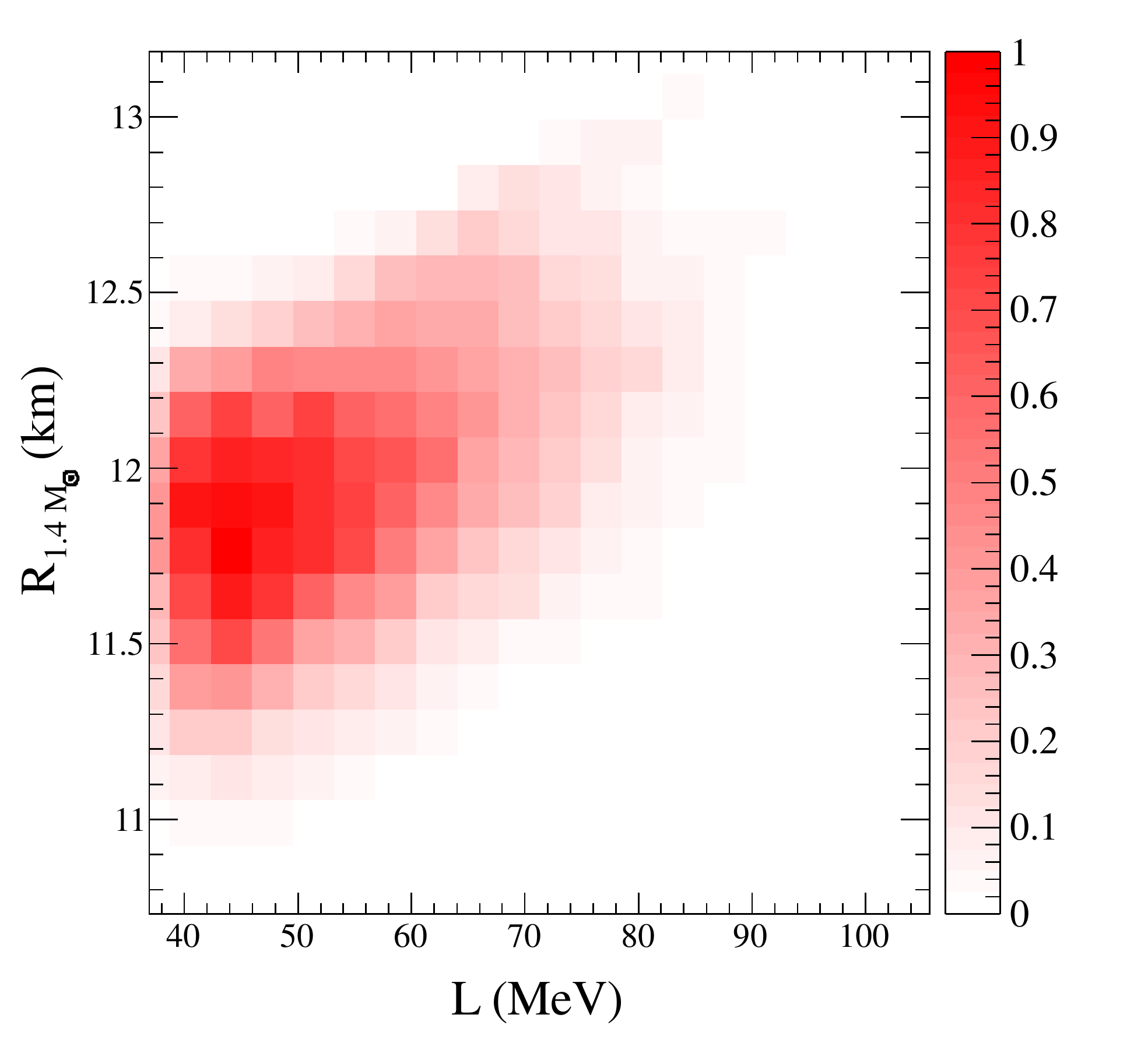}
\end{center}
\caption{Joint probability distribution of the radius of a 1.4 solar
mass neutron star and $L$ obtained using the data in Figs.~\ref{fig:pre_ns}
and~\ref{fig:qlmxb_ns}.
}
\label{fig:lvsr}
\end{figure}

This analysis also provides posterior probability distributions
for the EoS parameters. While we do not obtain significant
constraints on $a$ or $\alpha$, the mass and radius data do constrain
the parameters $b$ and $\beta$ (Fig.~\ref{fig:params}). While the
simple parametrization employed in this section cannot fully
describe the complexities of the nuclear three-body force, it does
make it clear that astrophysical data is beginning to rule out some
three-body forces which might otherwise be acceptable. 
Fig.~\ref{fig:lvsr} shows the joint distribution for the radius of a 1.4
$\mathrm{M}_{\odot}$ neutron star and $L$. The clear positive
correlation between these quantities demonstrates that the
observational data on neutron star radii constrains the value of $L$.
We find $32 < E_{\mathrm{sym}} < 34$ MeV and $43 < L < 52$ MeV to
within 68\% confidence~\cite{Steiner:2012}.

\section{Conclusions}
In this paper we report on our recent efforts to constrain 
the symmetry energy and its density dependence by combining microscopic 
calculations of neutron matter and observations of neutron star structure.
The EoS of pure neutron matter was calculated using   
Quantum Monte Carlo methods and a 
realistic nuclear Hamiltonian that includes the Argonne two-body
force and phenomenological three-body forces. Within the Illinois/Urbana 
model of three-neutron forces, we have extensively explored 
the role of the spin-isospin-independent short-range correlations 
that are mostly responsible for the behavior of the EoS at high density.
We have also studied the effect of the intermediate- and long-range 
contributions of the three-body force.
Within our model, we find that the uncertainty in the symmetry energy at saturation density 
dominates over other uncertainties associated with the short-distance structure of the 
three-neutron force. For the range of three nucleon forces employed we find a strong, 
nearly linear, correlation between the symmetry energy and its slope at saturation density.

An analysis of recent astrophysical measurements of several neutron star radii is shown 
to provide useful constrains on the slope of the symmetry energy.  To obtain agreement with 
neutron star radii extracted from this recent analysis we require a fairly repulsive three nucleon 
contribution to the energy.

\begin{acknowledgement}
We thank Steven C. Pieper for useful discussion regarding the content of 
this paper.
The work of S.G. and J.C. is supported by the U.S.~Department of Energy, Office of
Nuclear Physics, by the NUCLEI SciDAC program and by the LANL LDRD program.
The work of S.R. and A.W.S. is supported by DOE Grant No. DEFG02-00ER41132 and by the 
Topical Collaboration to study neutrinos and nucleosynthesis
in hot dense matter. 
The work of R.B.W. is supported by the US DOE Office of Nuclear Physics under
Contract No. DE-AC02-06CH11357.
The computing time has been provided by Los Alamos Open Supercomputing.
This research used also resources of the National Energy Research
Scientific Computing Center, which is supported by the Office of
Science of the U.S. Department of Energy under Contract No. 
DE-AC02-05CH11231.

\end{acknowledgement}

%\bibliographystyle{prsty}
%\bibliography{biblio} 

\end{document}